\documentclass[sigconf,screen]{acmart}

\setcopyright{acmcopyright}
\copyrightyear{2023}
\acmYear{2023}
\setcopyright{acmlicensed}
\acmConference[CHI '23]{Proceedings of the 2023 CHI Conference on Human Factors in Computing Systems}{April 23--28, 2023}{Hamburg, Germany}
\acmBooktitle{Proceedings of the 2023 CHI Conference on Human Factors in Computing Systems (CHI '23), April 23--28, 2023, Hamburg, Germany}
\acmPrice{15.00}
\acmDOI{10.1145/3544548.3580778}
\acmISBN{978-1-4503-9421-5/23/04}

\newcommand{\conditions}{\emph{Rest}, \emph{Twitter}, \emph{YouTube}, or \emph{TikTok}}

\usepackage{acmart-taps}
\usepackage{caption}
\usepackage{subcaption}

\usepackage{xcolor}
\definecolor{rest}{HTML}{00aa00}
\definecolor{twitter}{HTML}{1da1f2}
\definecolor{youtube}{HTML}{ff0000}
\definecolor{tiktok1}{HTML}{fe2c55}
\definecolor{tiktok2}{HTML}{25F4EE}

\usepackage{multirow}
\usepackage{tabularx}

\usepackage{dcolumn}
\newcolumntype{d}[1]{D{.}{.}{#1}}
\makeatletter
\newcolumntype{B}[3]{>{\boldmath\DC@{#1}{#2}{#3}}c<{\DC@end}}

\begin{document}

\title[Short-Form Videos Degrade Our Capacity to Retain Intentions]{Short-Form Videos Degrade Our Capacity to Retain Intentions: Effect of Context Switching On Prospective Memory}

\author{Francesco Chiossi}
\orcid{0000-0003-2987-7634}
\affiliation{
  \institution{LMU Munich}
  \city{Munich}
  \country{Germany}
}
\email{francesco.chiossi@ifi.lmu.de}

\author{Luke Haliburton}
\orcid{0000-0002-5654-2453}
\affiliation{
  \institution{LMU Munich}
  \city{Munich}
  \country{Germany}
}
\email{luke.haliburton@ifi.lmu.de}

\author{Changkun Ou}
\orcid{0000-0002-4595-7485}
\affiliation{
  \institution{LMU Munich}
  \city{Munich}
  \country{Germany}
}
\email{research@changkun.de}

\author{Andreas Butz}
\orcid{0000-0002-9007-9888}
\affiliation{
  \institution{LMU Munich}
  \city{Munich}
  \country{Germany}
}
\email{butz@ifi.lmu.de}

\author{Albrecht Schmidt}
\orcid{0000-0003-3890-1990}
\affiliation{%
  \institution{LMU Munich}
  \city{Munich}
  \postcode{80337}
  \country{Germany}}
\email{albrecht.schmidt@ifi.lmu.de}

\begin{abstract}
Social media platforms use short, highly engaging videos to catch users' attention. While the short-form video feeds popularized by TikTok are rapidly spreading to other platforms, we do not yet understand their impact on cognitive functions. We conducted a between-subjects experiment ($N=60$) investigating the impact of engaging with TikTok, Twitter, and YouTube while performing a Prospective Memory task (i.e., executing a previously planned action). The study required participants to remember intentions over interruptions. We found that the TikTok condition significantly degraded the users’ performance in this task. As none of the other conditions (Twitter, YouTube, no activity) had a similar effect, our results indicate that the combination of short videos and rapid context-switching impairs intention recall and execution.
We contribute a quantified understanding of the effect of social media feed format on Prospective Memory and outline consequences for media technology designers to not harm the users’ memory and wellbeing.
\end{abstract}


\maketitle

\section{Introduction}
Social media has exploded into ubiquity in recent years and is currently used by more than half the world's population~\cite{kemp_digital_2022}. Through social media feeds, people are frequently exposed to new information at an unprecedented rate. The information rate is also continuing to rise with the growing popularity of TikTok and other short video platforms, which displays a feed of brief, highly engaging videos. Although this high information density may be desired as it is beneficial for certain applications or appeals to users due to an instant gratification effect, it can also have serious negative consequences.  Digital information overload causes chronic stress~\cite{hefner_digital_2016} and social media use has been shown to have a detrimental impact on memory performance~\cite{schacter_constructive_2012}. We are still only beginning to uncover the breadth of the impact that social media can have on our psychological and cognitive functions.

TikTok's model of short, engaging videos that cause users to switch contexts rapidly is spreading across social media platforms (e.g., Instagram Reels~\cite{instagram_introducing_2020} and YouTube Shorts~\cite{jaffe_building_2020}). The widespread use of such video formats has certain known, negative consequences, including frequent disruptions, especially at the workplace~\cite{addas2018theorizing}. Online interruptions are the most common in the workplace~\cite{brumby2019interruptions} and are associated with increased workload, chronic stress, and mental fatigue~\cite{o1995timespace}. Specifically, this stream of video information continuously fills our mental buffer, which causes us to eliminate potentially useful information in favor of more superficial or irrelevant information provided by the social media feed. Past research in workplace contexts has demonstrated that context switching has a detrimental impact on cognitive functions and task performance~\cite{mark_cost_2008, spira_cost_2005}. However, the impact of context switching in social media is not well understood. \citet{zheng_influence_2021} found that watching short videos negatively impacts visual short-term memory, which suggests that we should better understand and characterize the impact of different social media feed formats on different cognitive functions. Prospective Memory (PM) is a fundamental cognitive function that describes our ability to remember to execute a planned action while doing an unrelated task~\cite{groot_prospective_2002}. PM is highly relevant because it enables productive activities, makes people effective in knowledge work, and accomplishing daily tasks (e.g., running errands or remembering to attend a meeting). Consequently, we aim to investigate the research question:
\begin{itemize}
     \item[\textbf{RQ}:] How do different social media feeds impact prospective memory?
 \end{itemize}

In this paper, we investigate how different social media feed interruptions impacts PM. We conducted a between-subjects study where $N=60$ participants simultaneously executed a lexical decision (LD) task and a PM task in two sessions with a 10-minute break in the middle. During the break, participants engaged in one of four activities, depending on their experimental condition: \conditions{}. Following the break, participants returned to the simultaneous LD and PM tasks to evaluate the impact on their PM. We hypothesized that \emph{TikTok} would have the most detrimental effect on PM because it is multi-modal, highly engaging, and exposes users to rapid changes in context.
Our results show that \emph{TikTok} significantly reduces PM performance relative to \emph{Rest}. On the other hand, we found that neither \emph{Twitter} nor \emph{YouTube} had any significant effect on performance. This paper contributes a quantified measure of the impact of three social media feed formats on PM, as well as an initial characterization of which social media feed format have the most significant impact on performance. Media technology designers need to understand the detrimental effect of exposing users to highly engaging rapid context switches, as poor PM performance can have drastic consequences for users in their daily lives.

\section{Related Work}
In this section we provide an overview of PM functioning and the impact of interruptions. Finally, we report the effects of social media distraction on cognitive functions related to PM that motivated our research.

\subsection{Prospective Memory}
PM is considered as ``the ability to remember to perform a previously planned action at a precise moment in time or following a specific event while one is engaged in performing another activity''~\cite{groot_prospective_2002}. It is a cognitive function that involves different cognitive processes, i.e.,  planning, attention, monitoring, and working memory. Individuals frequently become involved in multiple ongoing tasks after formulating an intention and, in most everyday contexts, cannot retain the intention in focal attention~\cite{cona2015neural}.

In the PM paradigm developed by Einstein-McDaniel \cite{einstein2005prospective}, participants are informed that if a particular target cue appears while doing an ongoing activity, e.g., a lexical decision task, they should execute a distinct action, such as pressing a specific key on the keyboard. Retrieving the delayed intentions requires a monitoring process mediated by bottom-up and top-down processes \cite{shelton2019multiprocess}. However, when we have to retain multiple intentions, we need to be in a preparatory state and actively monitor for the occurrence of target cues \cite{smith2007cost, einstein2005multiple}. Specifically,  monitoring requires allocating attentional resources for detecting the cue and memory resources to maintain and rehearse the intentions to execute \cite{guynn2003two}. Proper PM functioning is closely linked to productivity and safety~\cite{dismukes2012prospective} as it allows the execution of sequential steps such as programming \cite{sanjram2010attention}, office work~\cite{kvavilashvili2007time}, or taking or providing medication at the right time \cite{sanderson2015interruptions}.

Before we introduce the role of interruptions in PM, however, it is important to observe that, unlike any other memory task, in PM tasks, the recall of information does not occur as a result of an explicit request by someone~\cite{craik_effects_1996} (as for example in episodic memory tasks) but must be produced autonomously by the subject, in a self-initiated manner~\cite{mcdaniel_cue-focused_2004}.
Therefore, PM tasks do not only impose attentional demands but also memory demands for keeping the intention representation active and retrieving it~\cite{smith2014investigating, gordon2011structural}.

However, the cognitive demands of PM activities may not be limited to remembering intentions and monitoring cues. Therefore, this study focuses on another set of processes likely involved in PM performance, which has received little attention from PM researchers so far: the effect of temporary interruption of the ongoing and PM tasks.

\subsection{Interruptions in Prospective Memory}

As previously stated, PM paradigms encompass a dual-task nature in which participants are engaged in an ongoing activity while being asked to act upon perceiving a specific target cue.
However, outside laboratory settings, delays and interruptions frequently prohibit a person from carrying out an intended action after it is retrieved. Interruptions are pervasive in everyday life and at work, but there is a significant gap in our understanding of how such disruptions impair PM~\cite{mcdaniel2004delaying}. Specifically, in this work, we will focus on external interruptions. External sources of interruptions include face-to-face meetings~\cite{nees2015comparison}, instant messenger chats~\cite{gupta2013should}, workplace design features~\cite{oldham1991physical} and more. These examples of external interruptions also differ in the channel of interaction (i.e., direct or via technology~\cite{mcfarlane2002scope}), the sensory channels involved~\cite{lu2013supporting} or their information richness~\cite{chong2006interruptions}.

Multiple studies have demonstrated that after task interruption, task goals fade from memory, resulting in a long time to resume and complete the interrupted task, negatively impacting performance~\cite{altmann2014momentary, monk2008effects, adamczyk2004if}.
The detrimental behavioral impact of interruptions has been explained in terms of memory for goals theory~\cite{altmann2002memory}, focusing on memory-based deactivation of the interrupted task, or theory of attention residue~\cite{leroy2009so}, where the interruption retains attentional resources to some degree away from the user. Ultimately, the outcome is similar: interruptions have a negative impact on performance in the task at hand~\cite{leroy2018tasks, leroy2016effect}. Those two theories are intrinsically and functionally connected to how PM works. Interruptions implicitly require the involvement of PM processes as, after interruptions, we have to retrieve what we were engaged in and execute it~\cite{dodhia2009interruptions}.

When PM processes are interrupted, we are requested to resume the interrupted task~\cite{dodhia2009interruptions}. Then, we allocate residual attentional resources to monitoring prospective intentions~\cite{smith2003cost}, and to recalling and executing intentions again upon the appearance of the PM cue~\cite{cook2014role}. This repeated process with either expected memory fading or reduced attentional resources might induce a failure in these processes, resulting in forgetting about the halted work~\cite{mcdaniel2004delaying, rummel2017role}. In summary, interruptions are a cognitive burden that typically affect interrupted users' cognitive capacity to complete the interrupted task efficiently.

\subsection{Social Media as a Form of Distraction}

Distractions are caused by task-irrelevant stimuli that interrupt goal-directed behavior~\cite{clapp2012distinct}. Social media distraction is the process through which social media cues attract people's attention away from the task at hand (e.g., working). Social media interruptions differ significantly from workplace interruptions as they are more frequent (easily over 100 per day), more complex, and less predictable~\cite{zahmat2022mental}. This situation might regularly arise due to external (e.g., persistent notifications) or internal cues (e.g., unanswered messages)~\cite{wilmer2017smartphones}. Just as the above-mentioned interruptions, social media distractions also seem to affect memory performance, as shown in free recall tasks~\cite{sharifian2021daily, frein2013comes}. Those results are consistent with the idea that social media-induced distractions may harm memory functioning~\cite{fox2009distractions}. This detrimental effect might be explained with theories of memory for goals and attentional residue, as attentional disengagement interferes with memory encoding~\cite{hong2016attentional}.

Prior work in HCI has investigated methods for assisting users in using their phones more intentionally (e.g.,~\cite{terzimehic_mindphone_2022, hiniker_mytime_2016}), thereby avoiding being drawn in by the engaging strategies of social media designers. However, absentminded scrolling continues to be an issue~\cite{lupinacci_absentmindedly_2020}, motivating us to investigate the impact of social media feed designs.

More recently, social media apps started to employ different design strategies, e.g., video autoplay, pull-to-refresh, infinite scrolling, and recommendations, to maximize user engagement and attention capture~\cite{lukoff2021design}.
These designs create an immediate reward loop by showing content personalized to the user's subjective preference and interest~\cite{zeng2021research}, based on prior browsing histories and tagged video classification~\cite{bobadilla2013recommender, chen2019study}. The fast pace of switching between topics, which in the case of short form videos ranges from 15 to 60 seconds, and their often emotional content makes the social media distractor increase attentional disengagement~\cite{thorson1988effects, wirz2020prioritized}, and therefore impairs our capacity to timely and accurately resume the task at hand. However, there have only been a small number of studies on problematic short form video viewing habits, partly because short form video apps have only recently begun to proliferate.

The mechanisms underlying these effects are still not fully understood. However, recent work showed how a fast information rate might be linked to cognitive performance impairment in dual-task~\cite{uncapher2017media} and working memory settings~\cite{zheng_influence_2021, zahmat2022mental}. Dual-task settings and working memory are intrinsically related to PM, as PM encompasses multiple tasks to execute, and information to be kept updated in mind. Further research on the relationship between new forms of social media content and executive functioning is necessary, given their increasing pervasiveness and popularity in our daily life~\cite{mccoy2016digital, kemp_digital_2022}.
In this study, we investigated the effect of different social media feeds interruptions on executing previously planned intentions in a dual-task setting.

\section{Method}
We conducted a lab study using a between-subjects design with four \textsc{Interruption} conditions \conditions{}. The four conditions consist of three popular social media platforms with varying engagement styles and media formats as well as a control condition (\emph{Rest}) where participants are requested to not engage with any social media and do not perform any other actions.
We selected three social media platforms with large variation in feed format. \emph{TikTok} and \emph{Twitter} require users to rapidly switch contexts, while \emph{YouTube} generally consists of longer-format videos and therefore fewer context switches. The media format also varies between the three platforms. Both \emph{TikTok} and \emph{YouTube} are video-based while \emph{Twitter} is primarily text-based (with some photos). We did not include either Instagram or Facebook in this study, because their feeds are not sufficiently differentiable from \emph{TikTok} and \emph{Twitter} (i.e., they have a combination of videos, images, and text).

\subsection{Participants}
We recruited $N=60$ participants (35 female, 25 male, aged 19-34, $M=24.80$, $SD=3.40$) through a university mailing list and social media. All participants were fluent German speakers (C2 from CEFR \footnote{\url{https://www.coe.int/en/web/common-european-framework-reference-languages/level-descriptions}}), which was required for the LD task, and reported normal or corrected-to-normal vision with no history of any neurological or psychiatric disorders. The participants all had a high school education or higher, and reported a weekly average screen time of 2.04  hours ($SD=3.37$) in the \emph{Rest} condition, in the \emph{TikTok} condition 5.57 hours ($SD=2.25$), in the \emph{YouTube} condition of 6.75 hours ($SD=2.49$), and in the \emph{Twitter} condition of 5.51 hours ($SD=2.45$).
Moreover, we collected the screen time associated to the specific condition participants were allocated to. Specifically, participants allocated to the \emph{TikTok} condition spent an average of 1 hour and 46 minutes per week on the app ($SD=.81$). \emph{YouTube} participants showed an average screen time of 1 hours and 44 minutes ($SD=1.94$) on YouTube, while \emph{Twitter} participants spent an average of 55 minutes ($SD=.52$) on the Twitter social media app in the week before participation. Participants were randomly assigned to a condition containing a platform they use frequently in their daily life. We randomly assigned participants to either the app with the highest screen time in the previous week or to \emph{Rest}. We used this assignment method so that all participants have a personalized feed on the platform used in their condition. Rather than attempting to control the content on each feed, which would likely lead to some participants being uninterested in the content, we opted for this ecologically situated approach. The only exception to this allocation procedure was the YouTube condition, where we let participants choose one video out of ten options. This approach was chosen in order to control video duration but still allow participants to choose from a variety of contents e.g., education, music, and entertainment.

\subsection{Tasks}
We instructed participants to engage in a Lexical Decision (LD) task and a Prospective Memory (PM) task simultaneously, based on \citet{cona_theta_2020}. A large body of studies has demonstrated that this combination of tasks is suitable for monitoring prospective memory processes~\cite{einstein_multiple_2005, scullin_focalnonfocal_2010, cona_theta_2020}. Both tasks were conducted on a computer monitor (Acer Predator XB241YU 23.8 inch, 165 Hz, 2560 x 1440 pixels) with a mouse and keyboard. We covered all of the keys on the keyboard except for those used in the experiment (\textit{Q}, \textit{W}, \textit{E}, \textit{N}, \textit{M}, and \textit{Spacebar}). The experiment was created using PsychoPy~\cite{peirce2007psychopy}.

\begin{figure}[ht]
\centering
\includegraphics[width=\columnwidth]{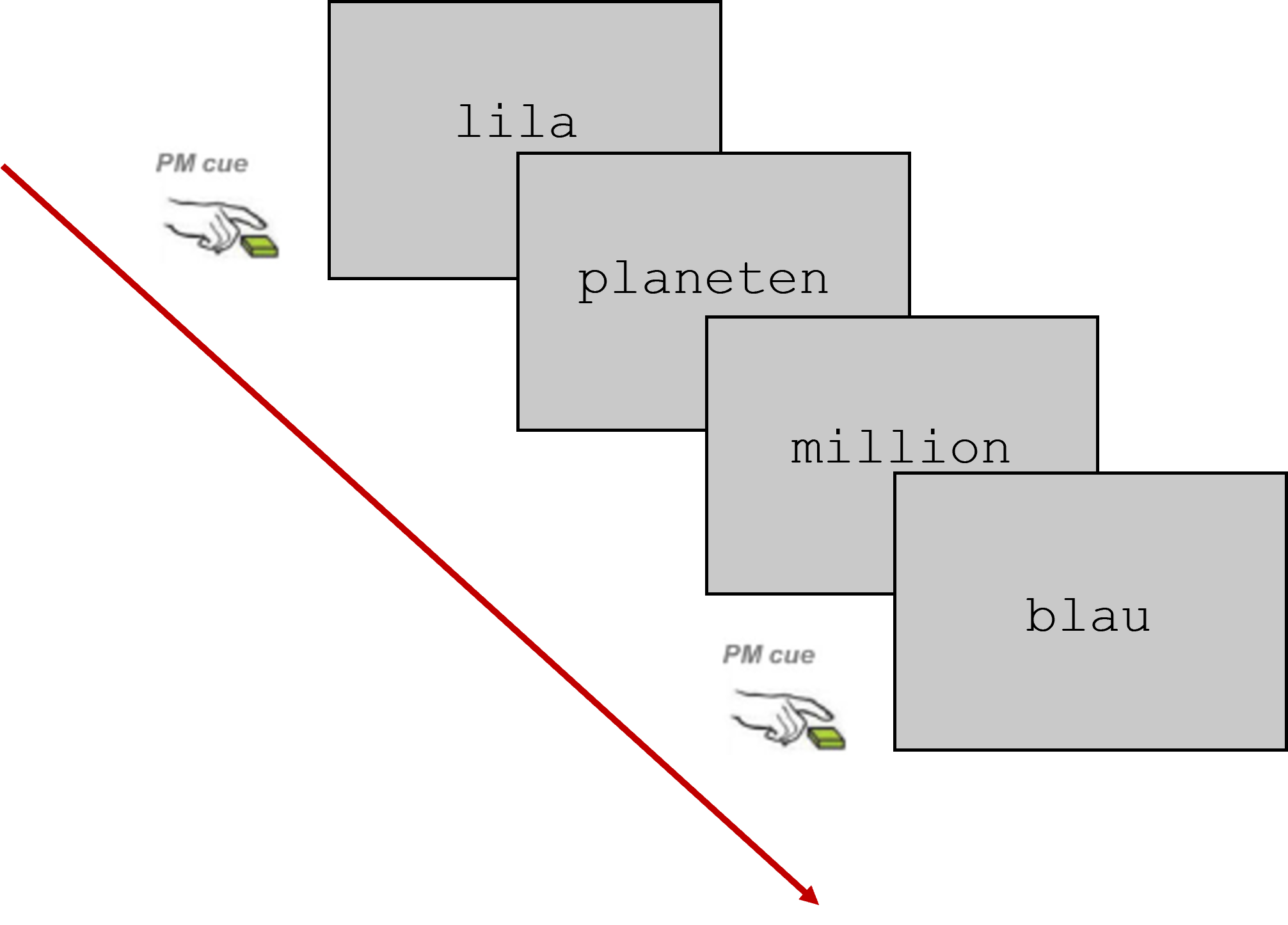}
\caption{PM paradigm. Participants performed a lexical decision task, i.e, performing a decision if one of the present words was either a word or not, and a PM task, where they were required to press a specific button in occurrence of a PM cue word ("blau", "lila" and "grün").}
\Description{PM paradigm. Participants performed a lexical decision task, i.e, performing a decision if one of the present words was either a word or not, and a PM task, where they were required to press a specific button in occurrence of a PM cue word ("blau", "lila" and "grün").}
 \label{fig:exp_paradigm}
\end{figure}

The tasks encompassed 160 LD trials and 16 PM trials (10\% of the LD trials) in two different blocks (Pre and Post Interruption), for a total of 320 LD  and 32 PM trials. Each experimental block started with a fixation cross (+) with a pseudorandom duration (1250, 1500, or 1750 ms) at the center of the screen. Then, a string of letters appeared as a stimulus for maximum 3000 ms. Inter-Stimulus-Interval was set to 1000 ms.
In the LD task, we displayed sequences of letters on the monitor and participants had to determine whether each sequence was a valid word or not. They pressed \textit{N} for words and \textit{M} for non-words with the index finger and the middle finger of the right hand, respectively. We counterbalanced the key-response mapping across participants. Word stimuli were extracted from the SUBTLEX-DE database~\cite{busch2022german}, with a word length ranging from six to eight letters. Psycholinguistic properties, such as the mean length and frequency, were matched across experimental sessions~\cite{brysbaert2011word}.
Non-words were pseudo-word stimuli created from the used words by changing one or two letters. Participants were required to respond to word stimuli as fast and accurately as possible. The PM task is depicted in \autoref{fig:exp_paradigm}.

In the PM task, participants were told to remember to carry out three different intentions linked to three different PM words. We instructed participants to press special keys if one of three keywords appeared on the screen instead of pressing the keys for the LD task. They were required to press \textit{Q} for 'Blau' (blue), \textit{W} for 'Lila' (purple) and \textit{E} for 'Grün' (green). Between a PM word, at least 10 LD trials appeared before the first PM cue and at least 8 LD trials between each pair of PM cues.  A practice block with just the LD task (five word/non-word trials) was given at the start of the experiment.
Participants used their personal mobile devices and headphones during the intermission tasks. We implemented all surveys using Google Forms.

\subsection{Interruption Conditions}
Each participant completed two rounds of the simultaneous LD and PM tasks with a 10 minute interruption in the middle, which varied according to their experimental condition. The four conditions for the interruption are as follows:

\vspace*{6pt}
\noindent{\textbf{Rest}}: Participants took a break with no input. We instructed them not to look at their phones or any other screen.

\vspace*{6pt}
\noindent{\textbf{Twitter}}: Participants were instructed to scroll through their own Twitter feed for the entire interruption. A Twitter feed consists of short texts with occasional photos. The feed switches contexts rapidly, but does not contain highly engaging video content.

\vspace*{6pt}
\noindent{\textbf{YouTube}}: We prepared a playlist of 10 minute YouTube videos in advance from a range of topics including entertainment and education (e.g., TED Talks). The participants were told to choose the video that was most interesting to them and watch it for the entire interruption. Although participants did not view their own personalized YouTube feeds, this method enabled us to control for the length of the videos, while still giving them some choice in content. The YouTube experience generally consists of longer-format videos with higher production value. By limiting the users to a single video, they do not switch contexts during the interruption.

\vspace*{6pt}
\noindent{\textbf{TikTok}}: Participants were instructed to watch videos on their own \emph{TikTok} feed for the entire interruption. A \emph{TikTok} feed consists of brief videos with sound. The feed is highly engaging and switches contexts rapidly.

\subsection{Procedure}\label{section:procedure}
\begin{figure*}[t]
\centering
\includegraphics[width=0.9\linewidth]{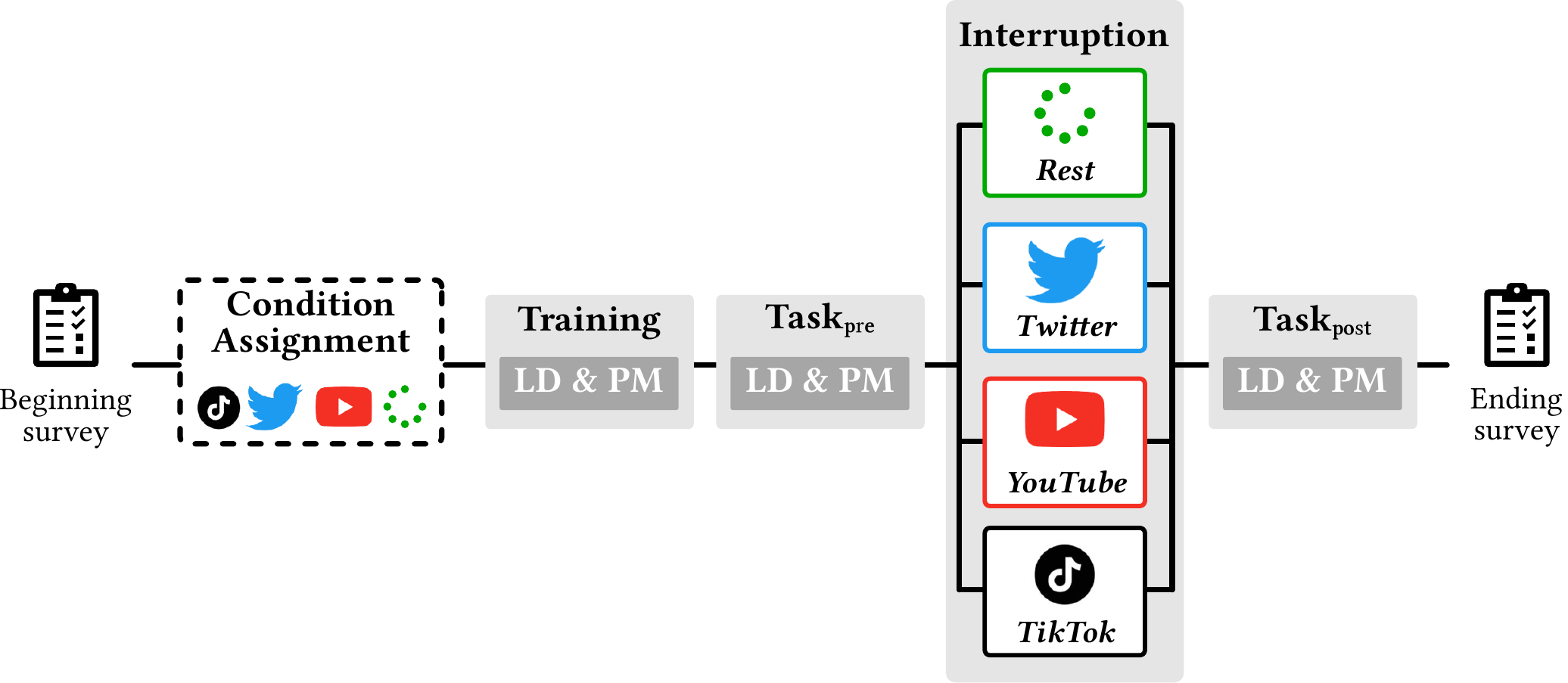}
\caption{Overview of the experiment timeline. Participants complete a Beginning Survey and are then assigned to one of 4 conditions. After a training phase, participants then perform LD and PM tasks in Task\textsubscript{Pre}, followed by an Interruption depending on their condition assignment (\conditions{}). Finally, the participants perform the LD and PM tasks again in Task\textsubscript{Post} and complete an Ending Survey.}
\Description{Overview of the experiment timeline. Participants complete a Beginning Survey and are then assigned to one of 4 conditions. After a training phase, participants then perform LD and PM tasks in Task\textsubscript{Pre}, followed by an Interruption depending on their condition assignment. Finally, the participants perform the LD and PM tasks again in Task\textsubscript{Post} and complete an Ending Survey.}
 \label{fig:timeline}
\end{figure*}

After introducing the study and obtaining informed consent, each participant proceeded through a beginning survey, four experimental steps, and finally an ending survey, as shown in \autoref{fig:timeline}.

In the beginning survey, we asked participants which social media platforms they use and the associated screen time. This information was used to assign participants to a study condition (\conditions{}).

First, in the  training stage, participants performed the combined LD and PM task for 10 trials. This stage ensured that the participants understood the requirements before conducting a measured task.
In Task\textsubscript{Pre}, participants performed the combined LD and PM tasks for 176 trials. Following Task\textsubscript{Pre}, the participants engaged in a break condition for 10 minutes. Depending on their assigned condition, the participants either engaged with a social media feed or took a break with no phone use.
After the break condition, the experimenter prompted participants to return to the computer and perform another 160 trials for the LD and 16 trials for the PM task (Task\textsubscript{Post}). Finally, participants completed an Ending Survey, reporting their overall screen time.

\subsection{Measures}
We measured behavioral data and collected responses to both a Beginning Survey and an Ending Survey. Each of these data sources require separate analysis methods. The questionnaires are included in the supplementary material.

In the beginning survey, we recorded which social media platforms each participant uses (``Which social media platform do you use between Twitter, TikTok and Youtube?''). Specifically, we asked participant to report how much time they spend in the three social media apps of interest. This information was used to assign participants to an appropriate condition, i.e., to the social media app they spent the most time with, and demographic information.  Regarding the behavioral performance, we measured accuracy and reaction times (RTs) for both the LD and PM tasks. Lastly, we analyzed the standard scales (BSMAS for social media addiction~\cite{monacis_social_2017}, and SUQ-A for absent-minded phone use~\cite{marty-dugas_relation_2018}) within the questionnaires according to their original documentation. We additionally collected Likert-scale responses on engagement and an Ending Survey for collecting screen time data (``Please enter your daily average screen time of the last week (search for Screen Time on iPhone or go to Digital Wellbeing on Android)'').

\subsection{Analyzing Performance Trade-Off}

We analyzed the reaction time distribution for the correct and error responses as two different distributions~\cite{brewer2011analyzing}.
To quantify the behavioral differences between correct and error responses, we adopted a \emph{Drift-Diffusion Model} (DDM)~\cite{ratcliff1978ddm}, which has been leveraged in cognitive science~\cite{horn2011can} and computer graphics~\cite{duinkharjav2022image} to model perceptual decision-making.
The DDM model also has been suggested to model \emph{choice} and \emph{non-choice} in LD and PM tasks~\cite{boywitt2012diffusion, cohen2017characterization, vinding2021volition}.
We chose this modeling approach as it is informative on different processing components relevant to PM. Traditional analysis approaches investigated RTs and task accuracy separately, profiting of only a subset of the available data, while the diffusion model is applied to the joint distribution of RT and accuracy data. DDM allows for the decomposition of RTs and accuracy into a set of latent parameters that represent underlying cognitive processes such as task load~\cite{ball2018importance}, evidence accumulation~\cite{smith2012prospective}, and decreased cognitive capacity~\cite{strickland2019evidence}.
The DDM assumes that decisions are made by a Wiener process that accumulates cognitive evidence over time from a starting point towards two response boundaries.

A DDM model includes 4 parameters: the drift rate ($\mu$), the decision bound ($B$), the non-decision time ($t$) and variance ($\sigma$) component. As an interpretation, the drift rate informs the speed and direction of information accumulation. The drift rate can be interpreted as a measure of subjective task difficulty: higher (absolute) drift rates indicate less demanding tasks. Second, the decision bound characterizes the time needed to make a decision, and with a larger decision bound, more effort is expected to form a decision. Here, smaller values imply shorter information uptake and increased erroneous responses. Third, the non-decision time captures the time spent for stimulus processing, but unrelated to the decision, such as perception of the target stimulus or execution of the response and stimulus encoding time. Lastly, the variance component models the uncertainty of a decision process, thus, an increased variance could result in more diverse reaction times needed to perform a decision when the PM stimulus is presented. Several studies have validated these parameters as sensitive to different experimental manipulations, lending credence to their validity~\cite{voss2008fast,wagenmakers2008diffusion,allen2014multitasking}.

\section{Results}

In this section, we first present results on behavioral accuracy using a Linear Mixed Model (LMM) approach. Second, we employ a Generalized Linear Mixed Model (GLMM) to investigate differences in the reaction times distributions. Finally, depending on normality, evaluated by the Shapiro-Wilk test~\cite{razali2011power}, we report two-way mixed ANOVA results for parameter analysis on the fitted DDM parameters, or ART ANOVAs~\cite{wobbrock2011art} for the non-parametric data. To analyze subjective responses collected from the engagement, SUQ-A, and BSMARS questionnaires, we performed a mixed ANOVA on the \textsc{Interruption} effect, with an additional Bayes factor ANOVA if the results were non-significant.

\begin{figure}[ht]
\centering
\begin{subfigure}[b]{0.49\textwidth}
\centering
\includegraphics[width=\textwidth]{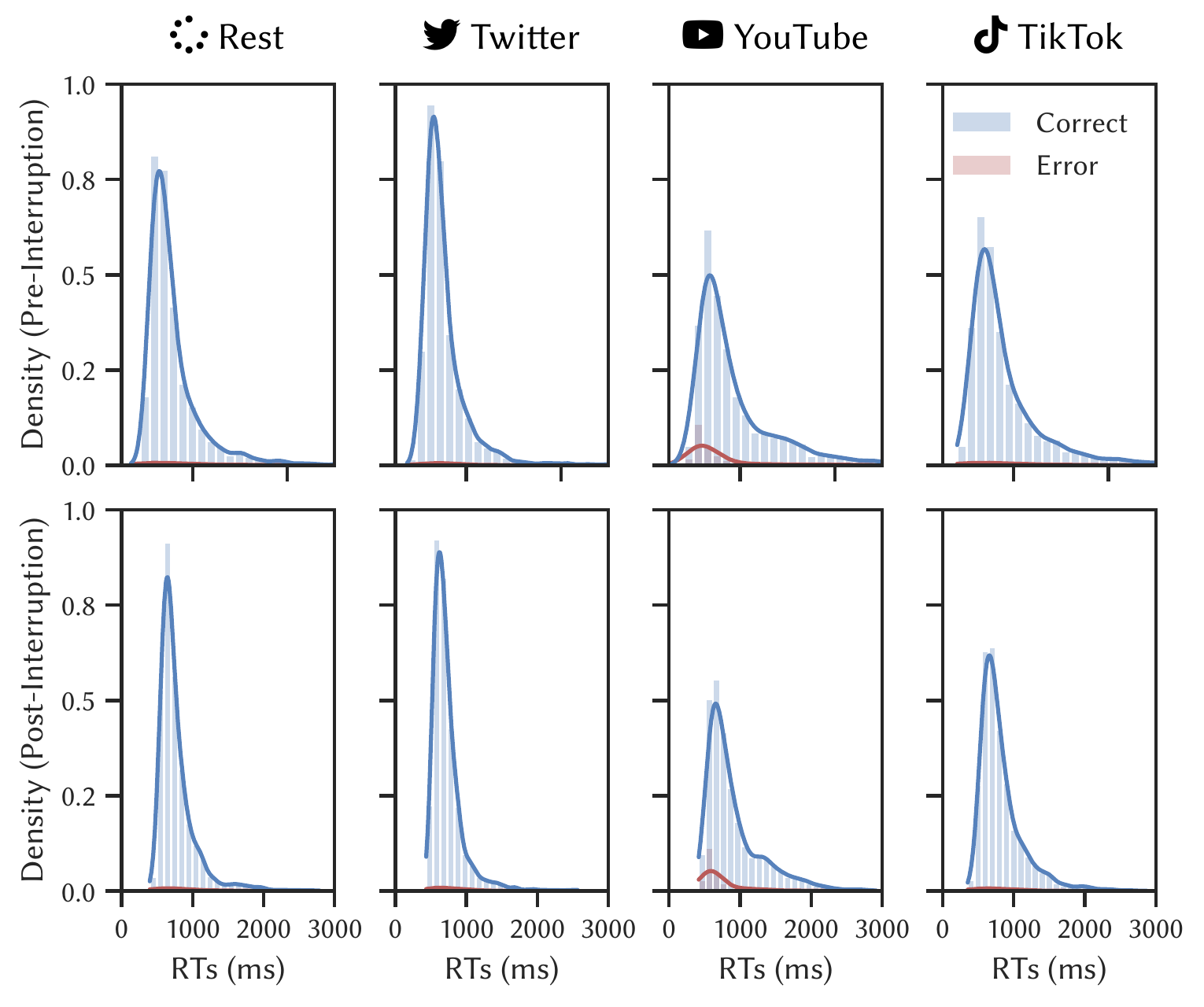}
\caption{LD Tasks}
\Description{LD Tasks}
\end{subfigure}
\begin{subfigure}[b]{0.49\textwidth}
\centering
\includegraphics[width=\textwidth]{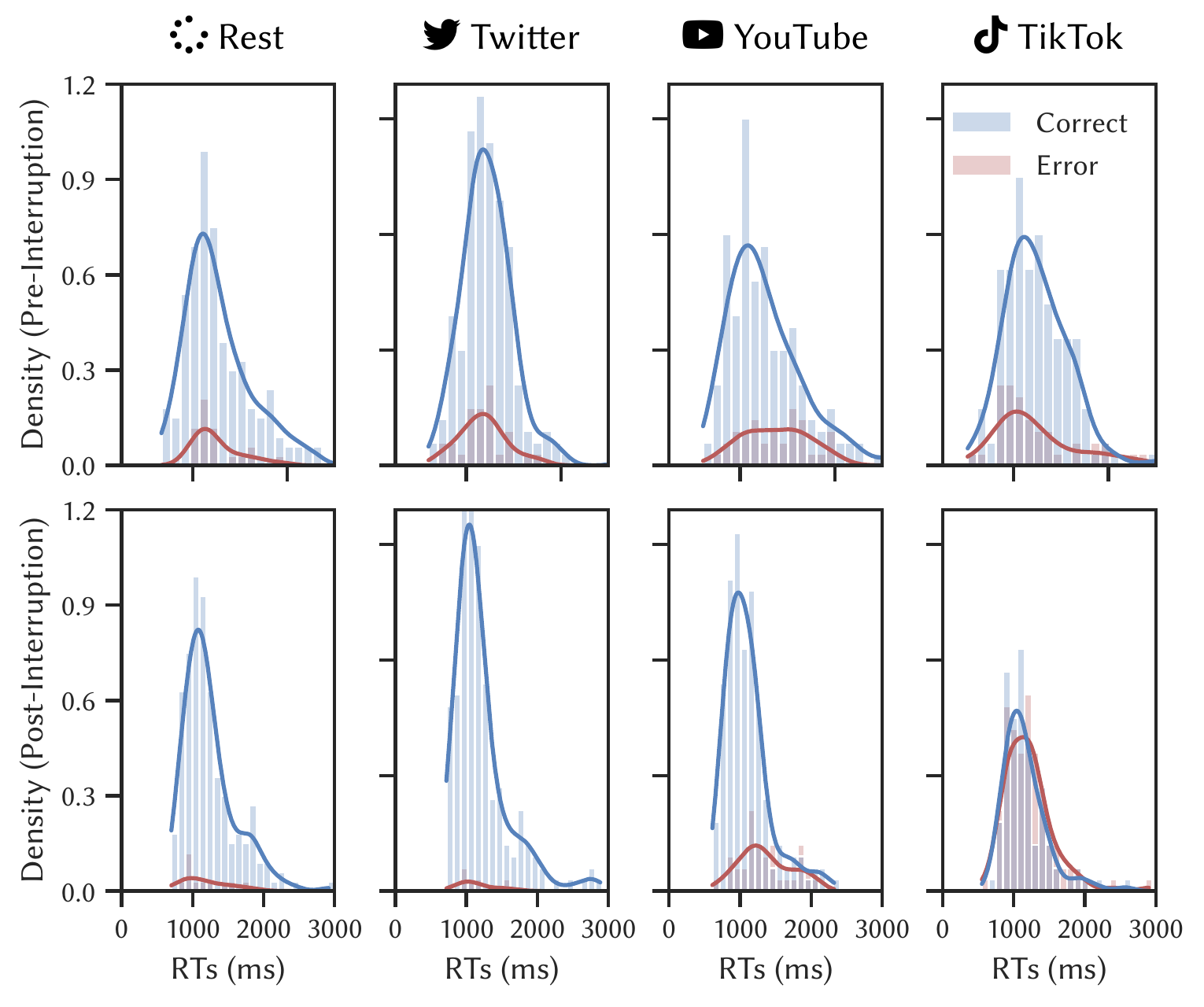}
\caption{PM Tasks}
\Description{LD Tasks}
\end{subfigure}
\caption{ An overview of reaction time distributions, separated by correct and error responses in the LD and PM Task.  The upper row (pre-interruption) shows reaction times before the interruption. The difference that are shown are due to the random assignment and are not linked to the condition. the lower row (post-interruption) shows the reaction time after the interruptions in which participants experienced different conditions.
The blue distributions show reaction time associated with correct responses and the red with error responses. A drastic increase in error responses can be seen in the \emph{TikTok} Post-Interruption PM trials (bottom right diagram).}
\Description{An overview of reaction time distributions, separated by correct and error responses in the LD and PM Task.  The upper row (pre-interruption) shows reaction times before the interruption. The difference that are shown are due to the random assignment and are not linked to the condition. the lower row (post-interruption) shows the reaction time after the interruptions in which participants experienced different conditions.
The blue distributions show reaction time associated with correct responses and the red with error responses. A drastic increase in error responses can be seen in the \emph{TikTok} Post-Interruption PM trials (bottom right diagram).}
\label{fig:rt-accuracy}
\end{figure}

\subsection{Behavioral results}

To give an overview of the reaction time distributions, we visualized the distribution for correct and error responses across different interruption conditions both in LD and PM tasks (see~\autoref{fig:rt-accuracy}). A drastic increase in error responses can already be seen in the \emph{TikTok} Post-Interruption PM trials (bottom right diagram). Below, we report our previously described statistical analysis of these distributions.

\begin{figure*}[ht]
    \centering
    \includegraphics[width=0.9\linewidth]{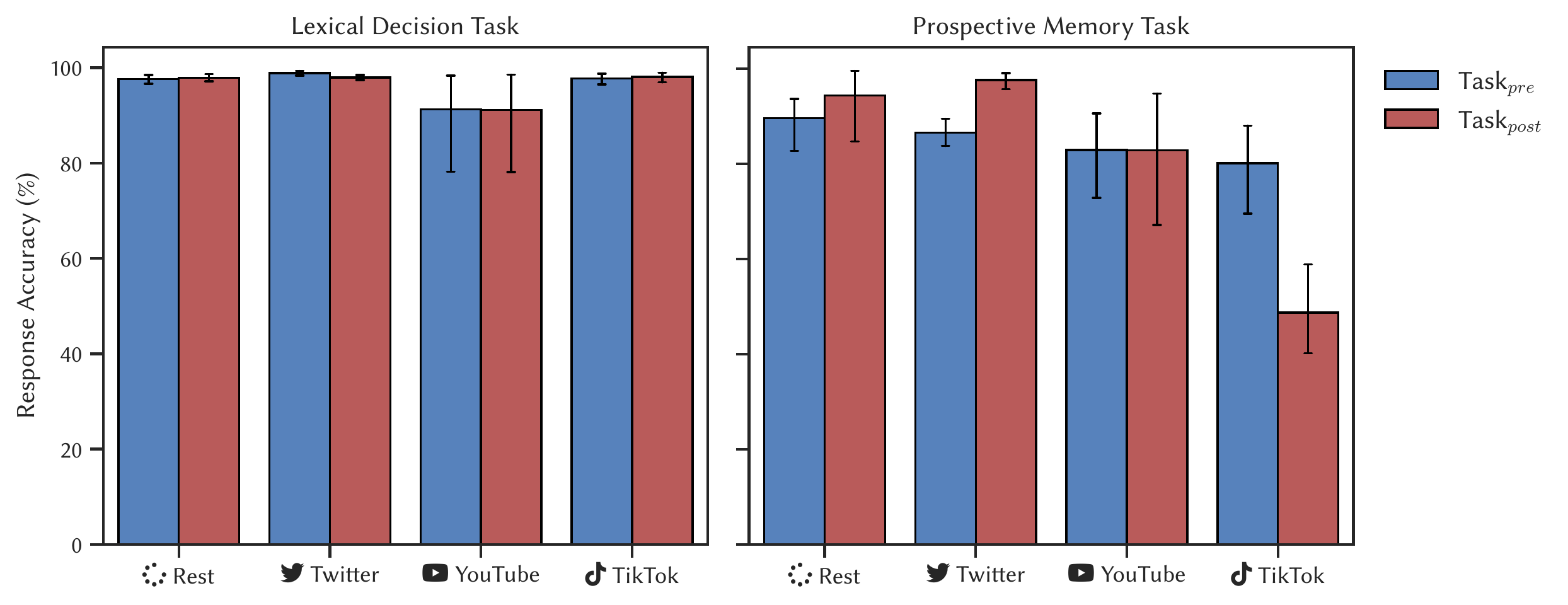}
    \caption{Comparison regarding response accuracy. The left figure visualizes pre- and post interruption in Lexical Decision task for different interruption conditions; the right figure visualizes for Prospective Memory task. The response accuracy barely changes (note the y-axis scale), whereas it drops dramatically for the PM task in the \emph{TikTok} interruption condition.}

    \Description{Comparison regarding response accuracy. The left figure visualizes pre- and post interruption in Lexical Decision task for different interruption conditions; the right figure visualizes for Prospective Memory task. The response accuracy barely changes (note the y-axis scale), whereas it drops dramatically for the PM task in the \emph{TikTok} interruption condition.}
    \label{fig:rt-accuracy-diff}
\end{figure*}

\subsubsection{Behavioral Accuracy}

As shown in Figure~\ref{fig:rt-accuracy-diff}, we inspected the response accuracy (total number of correct key presses divided by the total number of key presses) of participants\footnote{We further investigated trial by trial accuracy responses by means of a binomial regression accounting for per item and per participants effects. Results were completely consistent with the LMM model fit for PM and LD tasks. Results from this additional analysis are available in the supplementary material.}.

\paragraph{Lexical Decision Task}
We conducted an LMM ($R^2 = 0.67$) to predict accuracy with interruption (formula: \texttt{accuracy $\sim$ interrupt + (1|user\_id))} guided by REML and an nloptwrap optimizer and BIC criteria~\cite{peng2012model, barr2013random}. The model's intercept corresponding to \emph{TikTok} post interruption ($CI_{95\%}=[0.92, 1.04], t_{114} = 30.42, p < .001$). We found non-significance when comparing to other interruption conditions: \emph{Rest} ($\beta = -.001, CI_{95\%}=[-0.09, 0.09], t_{114} = -0.03, p = .975$), Twitter ($\beta = .005, CI_{95\%}=[-0.09, 0.10], t_{114} = 0.11, p = .913$), and \emph{YouTube} ($\beta = -0.07, CI_{95\%}=[-0.16, 0.02], t_{114} = -1.47, p = .144$) interruption conditions. Those results showed how any of the investigated interruption conditions impact the behavioral accuracy in the LD task.

\paragraph{Prospective Memory Task} Similarly, for PM task, we fitted an LMM model ($R^2 = 0.67$) to predict accuracy with interruption condition and measure (formula: \texttt{accuracy $\sim$ interrupt * measure + interrupt + measure + (1|user\_id)}). In the comparison with TikTok post ($CI_{95\%}=[0.40, 0.58], t_{110} = 10.94, p < .001$) interruption accuracy, we found significant and positive results for \emph{Rest} ($\beta = 0.46, CI_{95\%}=[0.33, 0.58], t_{110} = 7.25, p < .001$), \emph{Twitter} ($\beta = 0.49, CI_{95\%}=[0.36, 0.61], t_{110} = 7.77, p < .001$), \emph{YouTube} ($\beta = 0.34, CI_{95\%}=[0.22, 0.47], t_{110} = 5.42, p < .001$), as well as pre interruption condition ($\beta = 0.31, CI_{95\%}=[0.22, 0.41], t_{110} = 6.70, p < .001$). Therefore, we report that in the PM task, behavioral accuracy after interruption significantly dropped only in the \emph{TikTok} condition, while it remained stable across other conditions.

\paragraph{Behavioral accuracy Comparison between LD and PM tasks.}
Lastly, we fitted an LMM model ($R^2 = 0.38$) to predict accuracy with interrupt and task
 (formula: \texttt{accuracy $\sim$ interrupt * task + interrupt + task + (1|user\_id)}). We found only PM task on TikTok interruption condition is statistically significant and negative ($\beta = -0.28,  CI_{95\%}=[-0.39, -0.17], t_{230} = -5.01, p < .001$) comparing to LD task ($CI_{95\%}=[0.92, 1.04], t_{230} = 31.80, p < .001$) and other PM tasks (Twitter: $\beta = -0.005, CI_{95\%}=[-0.11, 0.10], t_{230} = -0.10, p = .923$; YouTube: $\beta = -0.03,  CI_{95\%}=[-0.13, 0.08], t_{230} = -0.46, p = .648$).

In sum, the abovementioned results show that in the \emph{TikTok} post-interruption trials, participants produced significantly more errors in the PM task, whereas the accuracy stayed robust in the LD task.

\subsubsection{Reaction Times}

\paragraph{Lexical Decision Task} In the LD task, we fitted a GLMM using REML and an nloptwrap optimizer on raw RTs considering \textsc{Interruption} as fixed effect and participant and stimulus item (word stimulus) as a random effects. We performed this analysis on RTs in both pre and post \textsc{Interruption}. We selected formula \texttt{rt $\sim$ interrupt * measure + interrupt + measure + (1|user\_id) + (1|stimulus)} for the GLMM with Gamma log link function~\cite{lo2015transform}, and guided by BIC criteria~\cite{peng2012model, barr2013random}. However, we did not report any significant results (See supplementary materials or \autoref{section:open}).
\paragraph{Prospective Memory Task} Using a similar approach, we conducted LMM with similar settings to PM task on raw RTs for both pre and post \textsc{Interruption}. Here, results mimicked the ones for LD as we did not report any significant difference (See supplementary materials or \autoref{section:open}). The distribution of RTs for both correct and erroneous responses is depicted in \autoref{fig:rt-accuracy}.

\subsubsection{DDM}

\aptLtoX[graphic=no,type=env]{
\begin{table*}[ht]%
\centering
\caption{An overview of  analyzed results using two-way ANOVAs on fitted DDM parameters. Significant results are highlighted in bold font. In terms of interruption, the results indicate significant differences in the drift rate $\mu$, variance $\sigma$, and decision bound $B$ in the PM task but no significance in the LD task. For pre- and post-interruption, the ANOVA showed a significant difference in variance $\sigma$ and non-decision time $t$ in the PM task but no significance in the LD task.}
\Description{An overview of  analyzed results using two-way ANOVAs on fitted DDM parameters. Significant results are highlighted in bold font. In terms of interruption, the results indicate significant differences in the drift rate $\mu$, variance $\sigma$, and decision bound $B$ in the PM task but no significance in the LD task. For pre- and post-interruption, the ANOVA showed a significant difference in variance $\sigma$ and non-decision time $t$ in the PM task but no significance in the LD task.}
\label{tab:anovas}
\begin{tabularx}{\linewidth}{X*{3}{d{2.0}d{2.0}d{1.3}d{1.3}d{1.3}}}
\toprule
& \multicolumn{5}{c}{\textbf{\textsc{Interruption}}}
& \multicolumn{5}{c}{\textbf{\textsc{Pre-Post}}}
& \multicolumn{5}{c}{\textbf{\textsc{Interruption $\times$ Pre-Post}}}  \\
\cmidrule(lr){2-6}\cmidrule(lr){7-11}\cmidrule(lr){12-16}
& \multicolumn{1}{c}{$f$} & \multicolumn{1}{c}{$df$} & \multicolumn{1}{c}{$F$} & \multicolumn{1}{c}{$p$} & \multicolumn{1}{c}{$\omega^2/\eta^2$}
& \multicolumn{1}{c}{$f$} & \multicolumn{1}{c}{$df$} & \multicolumn{1}{c}{$F$} & \multicolumn{1}{c}{$p$} & \multicolumn{1}{c}{$\omega^2/\eta^2$}
& \multicolumn{1}{c}{$f$} & \multicolumn{1}{c}{$df$} & \multicolumn{1}{c}{$F$} & \multicolumn{1}{c}{$p$} & \multicolumn{1}{c}{$\omega^2/\eta^2$} \\
\midrule
$\mu_{\text{PM}}$   & 3 & 56 & 4.078 & \multicolumn{1}{c}{\bf .011} & 0.133 & 1 & 56 & 2.420 & .125 & 0.024 & 3 & 56 & 6.466 & \multicolumn{1}{c}{\bf .001} & 0.215 \\
$\sigma_{\text{PM}}$ & 3 & 56 & 4.233 & \multicolumn{1}{c}{\bf .009} & 0.139 & 1 & 56 & 12.593 & \multicolumn{1}{c}{\bf .001} & 0.167 & 3 & 56 & 4.712 & \multicolumn{1}{c}{\bf .005} & 0.157 \\
$B_{\text{PM}}$ & 3 & 56 & 3.020 & \multicolumn{1}{c}{.037} & 0.092 & 1 & 56 & 0.025 & .874 & -0.017 & 3 & 56 & 2.385 & .079 & 0.065 \\
$t_{\text{PM}}$ & 3 & 56 & 1.695 & .179 & 0.034 & 1 & 56 & 15.851 & \multicolumn{1}{c}{\bf < .001} & 0.204 & 3 & 56 & 2.692 & .055 & 0.078 \\
\midrule
$\mu_{\text{LD}}$   & 3 & 56 & 2.520 & .067 & 0.119 & 1 & 56 & 1.033 & .314 & 0.018 & 3 & 56 & 0.785 & .508 & 0.040 \\
$\sigma_{\text{LD}}$ & 3 & 56 & 1.615 & .196 & 0.03 & 1 & 56 & 1.591 & .212 & 0.010 & 3 & 56 & 0.697 & .558 & -0.015 \\
$B_{\text{LD}}$ & 3 & 56 & 1.812 & .155 & 0.039 & 1 & 56 & 0.069 & .794 & -0.016 & 3 & 56 & 0.398 & .755 & -0.031 \\
$t_{\text{LD}}$ &3 & 56 & 1.517 & .220 & 0.025 & 1 & 56 & 0.607 & .439 & -0.007 & 3 & 56 & 0.636 & .595 & -0.019 \\
\bottomrule
\end{tabularx}
\end{table*}
}{\begin{table*}[ht]%
\centering
\caption{An overview of  analyzed results using two-way ANOVAs on fitted DDM parameters. Significant results are highlighted in bold font. In terms of interruption, the results indicate significant differences in the drift rate $\mu$, variance $\sigma$, and decision bound $B$ in the PM task but no significance in the LD task. For pre- and post-interruption, the ANOVA showed a significant difference in variance $\sigma$ and non-decision time $t$ in the PM task but no significance in the LD task.}
\Description{An overview of  analyzed results using two-way ANOVAs on fitted DDM parameters. Significant results are highlighted in bold font. In terms of interruption, the results indicate significant differences in the drift rate $\mu$, variance $\sigma$, and decision bound $B$ in the PM task but no significance in the LD task. For pre- and post-interruption, the ANOVA showed a significant difference in variance $\sigma$ and non-decision time $t$ in the PM task but no significance in the LD task.}
\label{tab:anovas}
\begin{tabularx}{\linewidth}{X*{3}{d{2.0}d{2.0}d{1.3}d{1.3}d{1.3}}}
\toprule
& \multicolumn{5}{c}{\textbf{\textsc{Interruption}}}
& \multicolumn{5}{c}{\textbf{\textsc{Pre-Post}}}
& \multicolumn{5}{c}{\textbf{\textsc{Interruption $\times$ Pre-Post}}}  \\
\cmidrule(lr){2-6}\cmidrule(lr){7-11}\cmidrule(lr){12-16}
& \multicolumn{1}{c}{$f$} & \multicolumn{1}{c}{$df$} & \multicolumn{1}{c}{$F$} & \multicolumn{1}{c}{$p$} & \multicolumn{1}{c}{$\omega^2/\eta^2$}
& \multicolumn{1}{c}{$f$} & \multicolumn{1}{c}{$df$} & \multicolumn{1}{c}{$F$} & \multicolumn{1}{c}{$p$} & \multicolumn{1}{c}{$\omega^2/\eta^2$}
& \multicolumn{1}{c}{$f$} & \multicolumn{1}{c}{$df$} & \multicolumn{1}{c}{$F$} & \multicolumn{1}{c}{$p$} & \multicolumn{1}{c}{$\omega^2/\eta^2$} \\
\midrule
$\mu_{\text{PM}}$   & 3 & 56 & 4.078 & \multicolumn{1}{B{.}{.}{1,3}}{.011} & 0.133 & 1 & 56 & 2.420 & .125 & 0.024 & 3 & 56 & 6.466 & \multicolumn{1}{B{.}{.}{1,3}}{.001} & 0.215 \\
$\sigma_{\text{PM}}$ & 3 & 56 & 4.233 & \multicolumn{1}{B{.}{.}{1,3}}{.009} & 0.139 & 1 & 56 & 12.593 & \multicolumn{1}{B{.}{.}{1,3}}{.001} & 0.167 & 3 & 56 & 4.712 & \multicolumn{1}{B{.}{.}{1,3}}{.005} & 0.157 \\
$B_{\text{PM}}$ & 3 & 56 & 3.020 & \multicolumn{1}{B{.}{.}{1,3}}{.037} & 0.092 & 1 & 56 & 0.025 & .874 & -0.017 & 3 & 56 & 2.385 & .079 & 0.065 \\
$t_{\text{PM}}$ & 3 & 56 & 1.695 & .179 & 0.034 & 1 & 56 & 15.851 & \multicolumn{1}{B{.}{.}{1,3}}{< .001} & 0.204 & 3 & 56 & 2.692 & .055 & 0.078 \\
\midrule
$\mu_{\text{LD}}$   & 3 & 56 & 2.520 & .067 & 0.119 & 1 & 56 & 1.033 & .314 & 0.018 & 3 & 56 & 0.785 & .508 & 0.040 \\
$\sigma_{\text{LD}}$ & 3 & 56 & 1.615 & .196 & 0.03 & 1 & 56 & 1.591 & .212 & 0.010 & 3 & 56 & 0.697 & .558 & -0.015 \\
$B_{\text{LD}}$ & 3 & 56 & 1.812 & .155 & 0.039 & 1 & 56 & 0.069 & .794 & -0.016 & 3 & 56 & 0.398 & .755 & -0.031 \\
$t_{\text{LD}}$ &3 & 56 & 1.517 & .220 & 0.025 & 1 & 56 & 0.607 & .439 & -0.007 & 3 & 56 & 0.636 & .595 & -0.019 \\
\bottomrule
\end{tabularx}
\end{table*}}

These observed behavioral accuracy results motivated us to further inspect participants' decision behavior in the LD and PM tasks using DDM for further interpretation. We used PyDDM~\cite{shinn2020flexible} to fit responses in the LD and PM tasks per participant. In sum, ANOVAs could not find any significance in all model parameters for the LD task. However, for the PM task, ANOVAs show significant differences in the drift rate, variance, and decision bound in different interruptions, and significant differences in the variance and non-decision time in pre- and post-interruptions. For detailed results, see~\autoref{tab:anovas}.

\begin{figure*}
    \centering
    \includegraphics[width=0.8\linewidth]{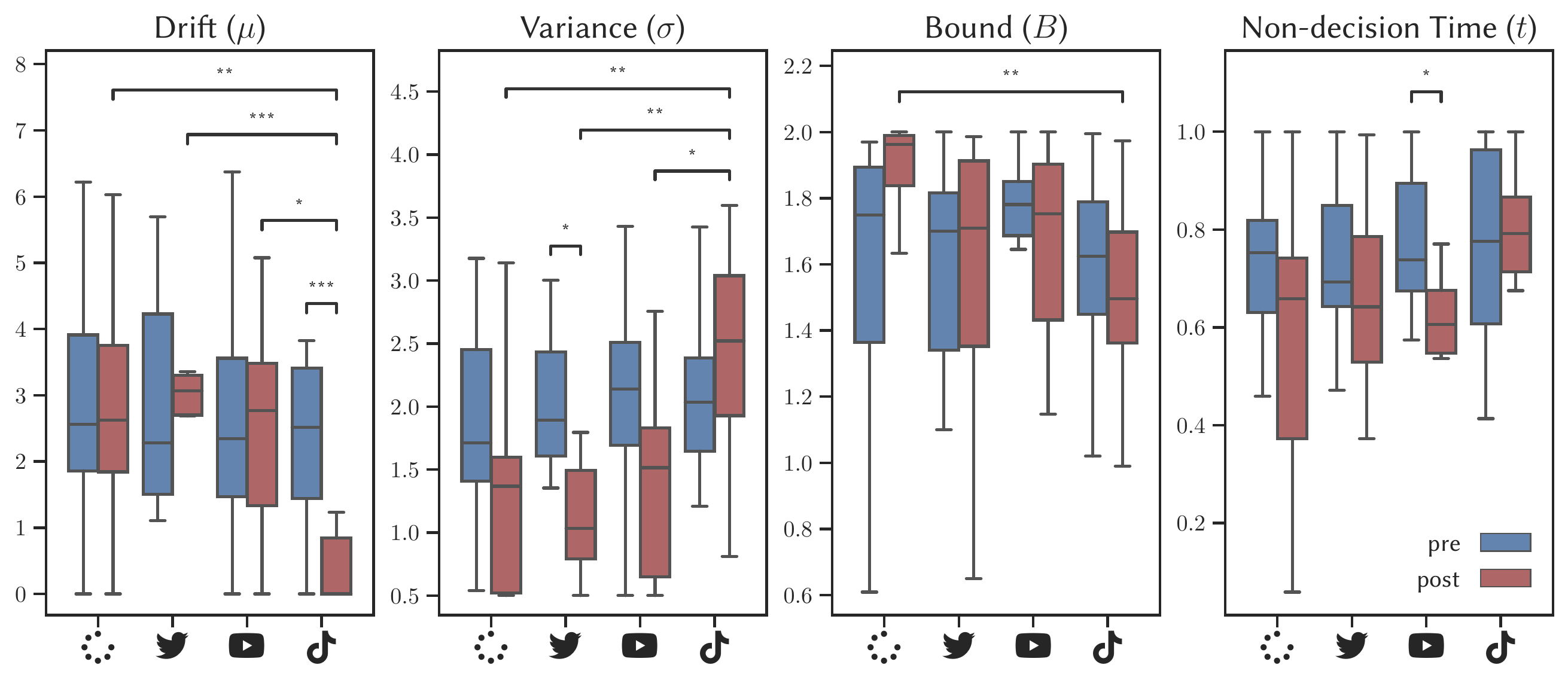}
    \caption{Post-hoc comparisons regarding fitted DDM parameters in the PM task. A larger drift rate $\mu$ means a stronger decision tendency toward correct responses, a larger variance $\sigma$ means higher uncertainty in a decision, and a larger bound $B$ requires more effort to form a decision. A larger non-decision time means less efficiency to start a decision.}
    \Description{Post-hoc comparisons regarding fitted DDM parameters in the PM task. A larger drift rate $\mu$ means a stronger decision tendency toward correct responses, a larger variance $\sigma$ means higher uncertainty in a decision, and a larger bound $B$ requires more effort to form a decision. A larger non-decision time means less efficiency to start a decision.}
    \label{fig:posthocddm}
\end{figure*}

Since LD tasks were non-significant, we only ran ART ANOVA post-hoc comparisons on PM tasks. \autoref{fig:posthocddm} shows the visualized results of these comparisons.
We found significant differences on $\mu_{\text{PM}}$ for \emph{TikTok} pre- and post-interruption contrasts comparison ($df=56, SE=8.456, t=-4.683, p<.001$), and significant differences on $\mu_{\text{PM}}$ across interruption conditions: \emph{TikTok} vs. \emph{Rest} ($df=91.45, SE=11.66, t=4.18, p=.002$), \emph{TikTok} vs. \emph{Twitter} ($df=11.66, SE=91.45, t=-4.67, p<.001$), \emph{TikTok} vs. \emph{YouTube} ($df=91.45, SE=11.66, t=-3.54, p=.014$).
For $\sigma_{\text{PM}}$, we also found significant differences for \emph{Twitter} ($df=56, SE=10.184, t=-3.283, p=.044$), and significant differences on $t_{\text{PM}}$ for \emph{YouTube} ($df=56, SE=8.470, t=-3.487, p=.014$). Across interruption conditions, we found significant differences on $\sigma_{\text{PM}}$: \emph{TikTok} vs. \emph{Rest} ($df=106.93, SE=11.52, t=-3.72, p=.009$), \emph{TikTok} vs. \emph{Twitter} ($df=106.93, SE=11.52, t=4.08, p=.002$), \emph{TikTok} vs. \emph{YouTube} ($df=106.93, SE=11.52, t=3.58, p=.014$), as well as \emph{TikTok} vs. \emph{Rest} ($df=112, SE=12.04, t=4.14, p=.002$) on decision bound $B$.

\begin{figure}
\centering
\begin{subfigure}[b]{0.49\textwidth}
\centering
\includegraphics[width=\textwidth]{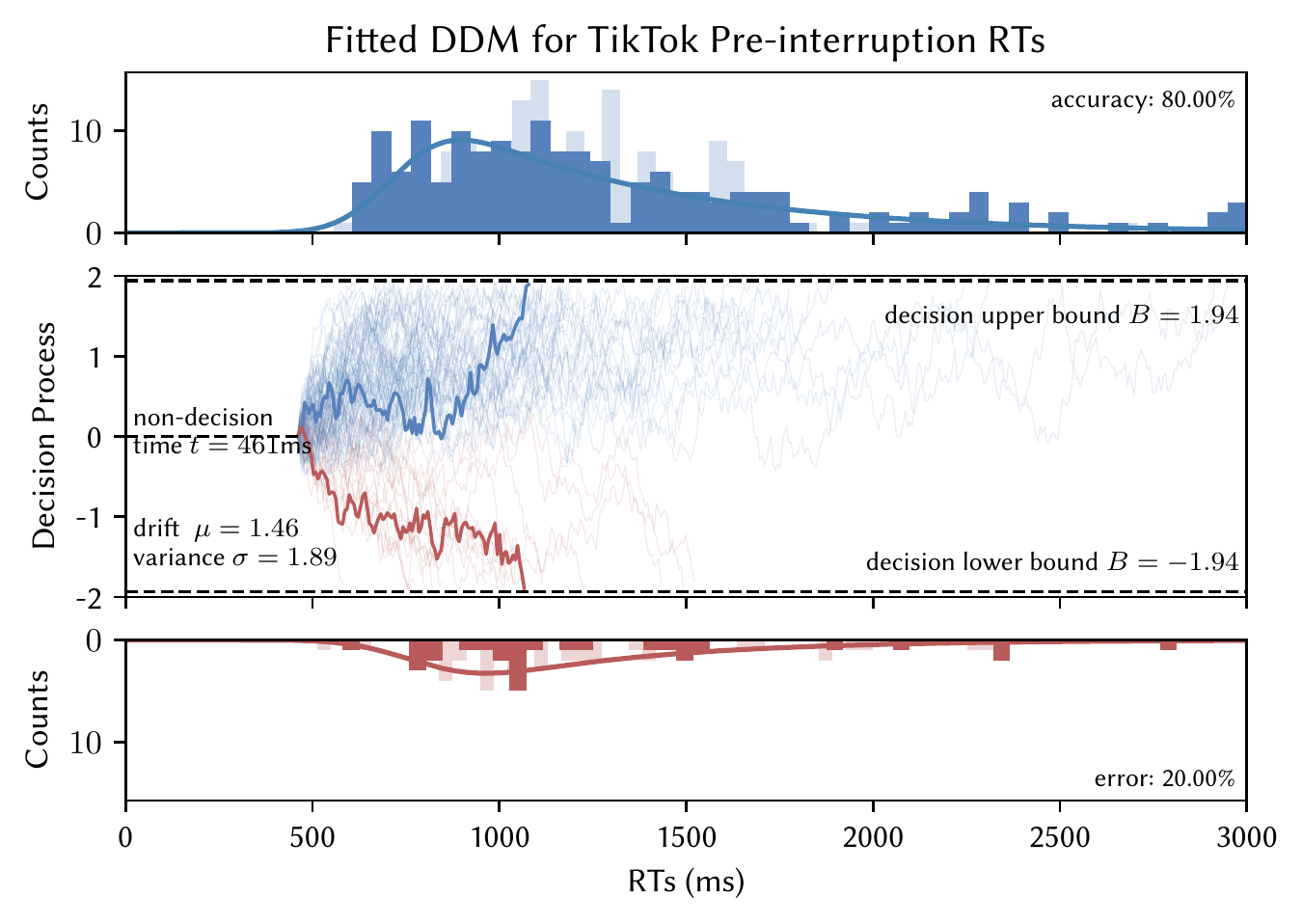}
\end{subfigure}
\begin{subfigure}[b]{0.49\textwidth}
\centering
\includegraphics[width=\textwidth]{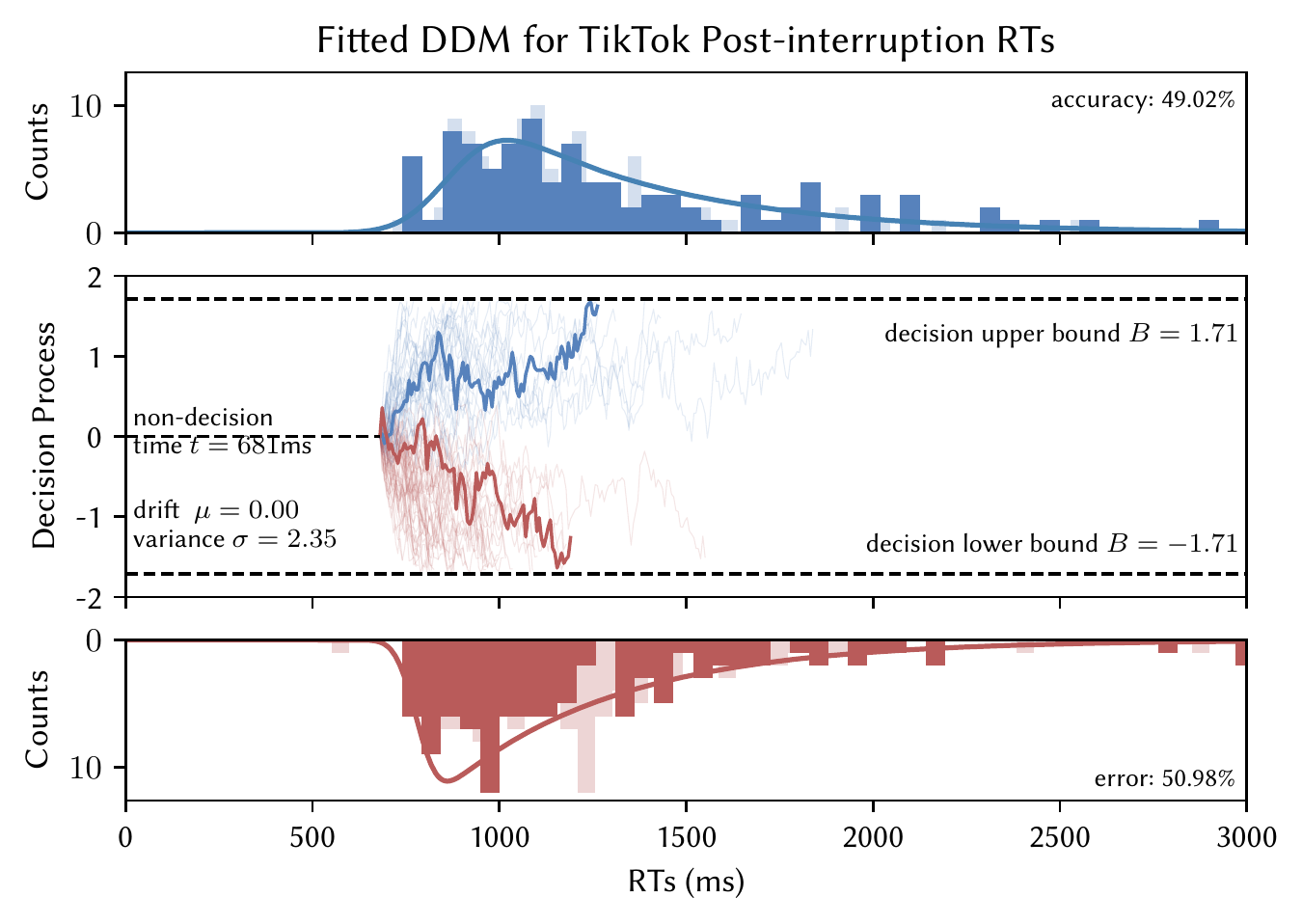}
\end{subfigure}
\vspace{-10px}
\caption{DDM visualizations for speed-accuracy tradeoff before (left) and after \emph{TikTok} interruption. The blue histograms on top show the distribution of correct responses, while the red histograms at the bottom show the error response distribution. Bold histograms represent correct and erroneous responses, and transparent histograms are DDM simulations. The middle figures are DDM simulated decision process. In \emph{TikTok}'s pre-interruption, the measured response accuracy was 80.00\%, with a drift rate $\mu=1.46$ towards correct responses. In the post- \emph{Tiktok} interruption, the measured response accuracy drops to 49.02\%, and the fitted DDM has a zero drift rate. \emph{TikTok}'s post-interruption has an equal tendency toward correct and error responses.}
\Description{DDM visualizations for speed-accuracy tradeoff before (left) and after \emph{TikTok} interruption. The blue histograms on top show the distribution of correct responses, while the red histograms at the bottom show the error response distribution. Bold histograms represent correct and erroneous responses, and transparent histograms are DDM simulations. The middle figures are DDM simulated decision process. In \emph{TikTok}'s pre-interruption, the measured response accuracy was 80.00\%, with a drift rate $\mu=1.46$ towards correct responses. In the post- \emph{Tiktok} interruption, the measured response accuracy drops to 49.02\%, and the fitted DDM has a zero drift rate. \emph{TikTok}'s post-interruption has an equal tendency toward correct and error responses.}
\label{fig:rt-ddm}
\end{figure}

For a closer look into the \emph{TikTok} condition, we visualized the DDM model (see~\autoref{fig:rt-ddm}) in pre- and post-interruptions for all 15 participants.
In pre-interruption, participants have a 20.00\% error rate, the fitted DDM model (total loss: 457.13) shows a drift rate $\mu=1.46$, variance $\sigma=1.89$, decision bound $B=1.94$, and non-decision time $t=461$ms.
In the post-interruption, participants have a total of 50.98\% error rate, the fitted DDM (total loss: 461.46) has a drift rate $\mu=0.000$, variance $\sigma= 2.35$, decision bound $B=1.71$, and non-decision time $t=681$ms.
These results show that our participants tended to give more correct responses ($\mu=1.46$) before the \emph{TikTok} interruption. However, after the interruption, participants had an equally probable decision tendency towards correct and error responses given a decision ($\mu=0.000$). Furthermore, the non-decision time increased from 461ms (pre) to 681ms (post), variance increased from 1.89 (pre) to 2.35 (post), and the decision bound was reduced from 1.94 (pre) to 1.71 (post).

\subsection{Subjective results}

We present subjective results separately for the reported engagement, SUQ-A, and BSMARS. Classical statistical inference was supplemented with Bayes Factors (BFs). This was done to establish the equivalence of interruptions on some of the dependent variables, which can be seen as a way of confirmatory testing of the Null-hypothesis~\cite{van2019bayes}.

\subsubsection{Engagement}

We performed a mixed one-way ANOVA on the subjectively reported engagement in each condition. We could not find any significant differences between \textsc{Interruption} conditions ($F_{3,56}=2.59, p=.062$). Hence, we further executed a Bayes factor ANOVA against the Null-hypothesis. The result specifically shows $BF_{01}=0.893$, which means that H1 is only 1.12 times more likely to occur than H0. This result implies insufficient evidence that engagement is different between conditions ~\cite{van2019bayes}.

\subsubsection{SUQ-A and BSMARS}

We performed a mixed one-way ANOVA on the reported SUQ-A values but could not find a significant difference ($F_{3, 56}= 2.267$,$p=.091$) across the different \textsc{Interruption} conditions. A subsequent Bayes factor analysis gives $BF_{01}=1.213$, meaning that H0 is only 1.213 times more likely than H1. Similarly, a mixed one-way ANOVA on the reported BSMARS could not find significant differences ($F_{3, 56}=1.065$, $p=.371$) across the \textsc{Interruption} conditions. A subsequent Bayes factor gives $BF_{01}=3.903$, meaning that H0 is 3.903 times more likely than H1~\cite{van2019bayes}.

\section{Discussion}

We evaluated the impact of different social media feed modalities on PM performance and RTs.
In our study, participants were instructed to simultaneously perform a LD and PM task with an interruption that varied according to the experimental condition (\conditions). We did this to understand whether and how different social media feed designs, with varying media modalities and context-switch frequencies, impact cognitive performance in terms of PM. We will first summarize our results, then relate our findings to the literature on how PM functions and on its relation with interruptions. We will conclude by summarizing possible consequences  for media technology designers and by highlighting future work in the field.

\subsection{Interpreting the Results}
\emph{TikTok} significantly degraded PM performance in terms of correct vs. erroneous responses. In fact, participants in the \emph{TikTok} condition were only slightly better than randomly guessing after the interruption. We can therefore conclude that the \emph{TikTok} condition had a significant negative impact on PM, while neither \emph{Twitter} nor \emph{YouTube} had any observable effect. We originally hypothesized that \emph{TikTok} would have a larger detrimental effect on PM than the other social media formats because it is highly engaging. However, we found no significant difference in subjective engagement scores between any conditions, so it appears that the effect is not caused by participants being more engaged in the \emph{TikTok} feed. Furthermore, we investigated if PM functioning might be impacted by video format or by rapid changes in context. However, as neither \emph{YouTube} nor \emph{Twitter} degraded PM performance, this result seems to point towards the detrimental impact of the combination of those two features, as shown in the \emph{TikTok} interruption. However, this result could be caused by other factors. In sum, we found that \emph{TikTok} significantly degrades PM performance, but further research is required to understand the exact mechanism underlying this effect.

Additionally, we quantified participants' social media addiction (BSMARS) and absentminded smartphone use (SUQ-A) to investigate whether the change in PM was influenced by variations in our sample populations. However, neither of these measures showed a significant difference across interruption conditions. Therefore, based on these three results, performance in the \emph{TikTok} condition was not affected by measured individual differences of the participants but rather by characteristics of the feed itself. It is theoretically possible that users of specific apps, such as \emph{TikTok}, might share specific characteristics or cognitive behaviors, such as a differential use of social media that might impact psychological and cognitive features at different levels \cite{sindermann2020predicting,montag2021psychology}. Although we found no effect from differences in SUQ-A or BSMARS, we cannot exclude that there might have been hidden mediatory variables, such as daily negative affect \cite{sharifian2021daily}. Therefore, future work is necessary to compare the interruption effect on users unfamiliar with a specific app. This would further investigate the relationship between PM and specific social media formats and help uncover the mechanism driving social media-based PM degradation.

\subsection{Impact of Social Media on PM}

Our study exposed participants to different social media feeds format. Our main objective was to determine whether social media interruptions impact PM retrieval and monitoring to identify criticalities for technology design and manage the adverse effects of social media interruptions.
Short-form videos represent multimodal and emotional stimuli, whose content is often tailored to their consumer~\cite{davidson2010youtube}. They are salient interruptions that quickly divert attention. Thus, when a PM task is interrupted, people might not have enough time or be too distracted to resume the intention explicitly. This is because such dynamic visual and auditory features require attention focus for effective information processing during video watching~\cite{min2020multimodal}. However, such attentional resources demanded by the video format have been shown to impact the participant's capacity to either identify PM cues for triggering intention execution or keep PM intentions active in mind.

In the PM task, we reported three interesting results that corroborate our interpretation of the detrimental effect of short-term videos.
First, the PM cue detection accuracy decreased by almost 40\% after users were interrupted by short-form videos. After the \emph{TikTok} interruption, participants tried to retrieve the intention, but this goal might be fleeting because of the attentional demands of the interruption. Even if participants were aware of the need to retrieve the intention, they were not able to associate the PM cue word with the associated button to press.
This interpretation is in line with previous research showing that dividing attention impaired PM performance~\cite{einstein1997aging, mcgann2002conceptual} and visual working memory performance~\cite{zheng_influence_2021}. Moreover, as the interruption recruited attentional resources, the interruption duration did not leave enough time to retrieve and successfully resume the intention. Participants were vulnerable to the interruption and not ready to resume the task, as they were cognitively engaged with the interruption. This result is in line with earlier research showing that participants who retrieve an intention sometimes forget to execute it even after shorter interruptions~\cite{mcdaniel2003aging, dodhia2009interruptions}.

Second, the DDM parameter analysis allowed us to examine the cost and interference effects of the interruption. DDM integrates two behavioral features, which are usually in a compensatory relationship (i.e., accuracy and response speed), into psychologically meaningful process parameters~\cite{ratcliff2013parameter}. The first parameter extracted with this approach, the drift rate, mimicked the results we obtained with behavioral accuracy post-interruption. Drift rate represents processing efficiency and therefore, it is selectively influenced by the task demands on participants~\cite{rummel2013performance}. In our study, short-form videos increased the memory demand on participants, making it harder to process each alternative, leading to reduced evidence accumulation rates for making a decision. The TikTok interruption absorbed resources that would otherwise be allocated to the PM task, thereby slowing processing efficiency~\cite{smith2008connecting,smith2010costs}.
This interpretation is specifically supported by the fact that we did not find any significant difference when analyzing the parameters from the DDM in the LD task.

Similarly to the drift rate, the variance parameter also showed the same results as the drift rate. Participants had increased uncertainty about which intentions to execute, i.e., which associated button to press. This can be seen in the light of reduced attentional capacities. Theories of attention associate resource allocation to preparatory responses and increased processing speed~\cite{langner2010mental}. The time needed for accumulating evidence and making a decision is influenced by the attention directed towards it. Thus, short-form video consumption competed for such resources and resulted in decreased evidence accumulation in the PM task~\cite{boag2019strategic}.
The drift rate and variance results align with the overlapping attentional and memory demands that affect accuracy and RT variability via longer RTs during PM attentional lapses~\cite{ihle2017prospective}.

Finally, we found decreased decision bound in the short-form video compared to the control condition (\emph{Rest}). This implies that participants chose a more liberal response criterion, and therefore, they accepted less evidence to make a response after the \emph{TikTok} interruption. As a result, participants failed to retrieve the intention and, therefore, made a less informed decision, impairing their accuracy. They were not ready, given their excessive memory load, to monitor the PM cues and make a conscious decision~\cite{boywitt2012diffusion}.
Taken together, those results point towards how PM is vulnerable to context-switching, specifically in the form of engaging short-form videos.

\subsection{Consequences for Media Technology Designers}

Our results have  consequences in two major areas: mitigating the adverse effects of short-form videos on PM and intentionally exploiting the associated PM degradation.

\subsubsection{Mitigate the adverse effects of short-form videos on PM}
We show that interruptions with short-form video (such as but not limited to \emph{TikTok}) significantly degrade PM post-interruption, which means that a user is likely to have degraded performance when they return to their primary task. Prospective memory is a crucial aspect of daily life and is susceptible to interruption. Therefore, we argue that a social media feed that negatively impacts cognitive performance in the real world is generally not desired and could be classified as a \textit{Dark Pattern}. \citet{gray_dark_2018} outlined five strategies for dark patterns: nagging, obstruction, sneaking, interface interference, and forced action. Degrading cognitive performance in the real world does not fall into any of these categories but perhaps deserves its own classification. Recently, multiple platforms are introducing features inspired by the \emph{TikTok} feed (e.g., Instagram Reels and YouTube shorts). By increasing the use of this dark pattern across the social media landscape, social media designers might impact PM performance across a broader audience. Additional research is required to understand how to combat this effect, but we can suggest some logical first approaches. Past work in  HCI has used digital reminders~\cite{wang_exploring_2014} and other memory aids~\cite{schulze_memos_2003} to improve PM performance. Moreover, digital reminder systems have been shown to support rehearsal for highly-specified intentions~\cite{brewer2017how}, such as remembering to purchase a specific item in the supermarket. Similar reminder-based approaches could help users remember tasks after a social media interruption in a productivity context. Alternately, proactive reminder systems~\cite{yorke2009like} or Digital Self-Control Tools \cite{lyngs2019self} can anticipate the social media feed engagement and suggest positive and healthier interruptions~\cite{sasangohar2012not}. This would, first, support the overall efficiency of digital reminders for task management and, second, overcome challenges in support of PM as intentions are generally triggered by either specific events or times~\cite{jager2008time}. However, in line with past work on mindful smartphone use~\cite{terzimehic_mindphone_2022, lukoff_what_2018}, we highlight the fundamental necessity for users to maintain agency and be permitted to use their smartphones and social media feeds in a way that aligns with their own needs and values.

Finally, since we only observed PM degradation in the \emph{TikTok} condition, it follows that interspersing short-form video feeds with other media formats may mitigate this effect. Investigating interventions to prevent the disruption and generally enhance prospective memory is important from a practical standpoint~\cite{mcdaniel_prospective_2007} and may potentially improve cognitive performance and well-being for a generation of internet users. %

\subsubsection{Leverage the impact of short-form videos to intentionally make users forget information}
Aside from avoiding adverse effects, technology designers could exploit  PM degradation by intentionally engaging short-form videos as interruptions in strategic situations. Such interruptions could be employed to help a user forget information. For example, this effect could be used in a video game where the designer wants to challenge the player by disrupting their ability to remember a task they need to accomplish. Distraction is a recommended game design technique to increase immersion or the impact of surprising scenes~\cite{rogers_level_2014}, but it could also increase difficulty. To operationalize this, a game designer could strategically insert short cutscenes to distract a player from a task they are supposed to remember to perform. This intentional distraction could increase difficulty while entertaining the user with engaging scenes.

This approach could also be employed therapeutically to help a user forget about unhappy memories. Distraction through media, such as video games, has been effectively used for mental health recovery applications~\cite{colder_carras_connection_2018}. In practice, this could be combined with an affective computing paradigm that monitors when a user is unhappy and intervenes with short videos to disrupt their memories. However, additional research is required to determine whether short-form videos could be used for such applications.

Finally, designers could employ short-form videos to help users transfer their focus to a new task. Research on task-switching shows that the initial task interferes with the second task because users maintain residual attention on features that are no longer relevant~\cite{richter_memory_2012}. This could be implemented by triggering a series of short-form videos any time a user switches their focus to a new task, which may reduce the residual attention on the previous task and thereby reduce the time required to fully focus on the new task. There is a need for additional future work to determine whether short-form videos could be used to diminish the adverse effects of residual attention.

\subsection{Limitations \& Future Work}

Our results must be viewed in light of certain limitations, which we reflect on below.

We did not evaluate every social media platform, however, we chose three of the currently most widespread platforms, which largely vary in their feed format. Other major social media platforms are either very similar to platforms we have already chosen (e.g., Mastodon shares nearly all feed features with Twitter), or consist of a combination of features from the platforms we selected (e.g., Instagram and Facebook both have `Reels', which are user-generated short-format videos similar to TikTok, Facebook feeds also contain longer videos, similar to YouTube, and both have text and image posts of variable lengths, similar to Twitter).

Second, users did not browse their own feeds in the \emph{YouTube} condition. Rather, they freely chose a video from a pre-defined list of options. This approach was chosen to control for video duration and equalize the interruption length, as different interruption length has shown to have differential effects on information encoding and retrieval~\cite{morgan2009improving,altmann2019integrating,monk2008effect}, thus compromising between time demands and content preference. Future work will address how the effect of long format videos can impact PM with content chosen by the users and for different duration.

Additionally, our situated experimental design presented confounding variables. We chose three popular social media feeds that varied in terms of media content. \emph{Twitter} mostly relies on text content with sporadic images, \emph{Youtube} on video content of medium duration ($\sim$ 11.7 minutes)\footnote{\url{https://www.statista.com/statistics/1026923/youtube-video-category-average-length/}}, while \emph{TikTok} also employs video content but with a short duration (15 seconds -- 3 minutes). The social media feeds also differed in the pace of the presented content, i.e., context switching and media format. In our study, we did not attempt to quantify how each feed characteristic (number of context switches and media format) individually contributes to PM degradation. Therefore, we did not employ a condition that combines text content as \emph{Twitter} and no context switching as \emph{YouTube}, e.g., a 10-minutes reading article interruption. However, media modality has shown to differentially impact memory processing depending on textual or visual stimulation~\cite{lang1999something}, or when such content is combined~\cite{zhang2022heavy}, interfering with automatic and control processing \cite{powell2019framing}.
Thus, now that our research has identified that different social media feeds have different impacts on PM, this should motivate future work to investigate this effect on a more fine-grained level. We propose a future experiment that systematically varies the pace of context switching as well as the type of media modality.

Finally, the lexical decision task is an established ongoing task when investigating PM~\cite{einstein2005prospective}, but it may not be the most ecologically valid choice. Therefore, future work should aim to extend our results to real-world tasks, which might bring to light additional effects that remained hidden in our laboratory setting. One potential direction in such a more realistic setting is the design of reminders that could counteract the detrimental effect of interruptions, which has successfully been shown in other contexts~\cite{Kern2010GazemarksGV}.
Thus, it could be desirable to compare the effect of encoding reminders with a no-reminder condition and a pause condition that allows participants ample time to encode. Some reminder cues, but not others, have been shown to improve PM performance in PM paradigms~ \cite{guynn_prospective_1998}.

\section{Conclusion}
In this paper, we investigated the impact of social media on prospective memory. We conducted a between-subjects study with 60 participants comparing three social media feeds (\emph{TikTok}, \emph{Twitter}, and \emph{YouTube}) and a \emph{Rest} condition as a control. Our results show that short video streams such as \emph{TikTok} have a significant detrimental impact on prospective memory performance. Specifically, users showed a worsened speed/accuracy trade-offs as compared to all other experimental conditions. The other platforms do not significantly affect performance. Interestingly, social media addiction, absent-minded phone, and perceived engagement did not have any relationship or influenced accuracy across conditions. This allowed us to disentangle the effect of individual differences from the detrimental one of short-form videos on PM. We contribute an empirical understanding of the impact of different social media feeds on PM and discuss consequences for technology designers to create engaging experiences without negatively impacting users.

\section{Open Science}
\label{section:open}
We encourage readers to reproduce and extend our results and analysis methods. Therefore our experimental setup, collected datasets, and analysis scripts are available on Github \footnote{\url{https://github.com/mimuc/media-prospective-memory}}.

\begin{acks}
We would like to thank Martina Gluderer for her support during data collection and the submission of this work. Francesco Chiossi was supported by the Deutsche Forschungsgemeinschaft (DFG, German Research Foundation), Project-ID 251654672-TRR 161. Luke Haliburton was supported by the Bavarian Research Alliance association ForDigitHealth.
\end{acks}

\balance
\bibliographystyle{ACM-Reference-Format}
\bibliography{references}


\begin{thebibliography}{125}

%%% the Web address.

\ifx \showCODEN    \undefined \def \showCODEN     #1{\unskip}     \fi
\ifx \showDOI      \undefined \def \showDOI       #1{#1}\fi
\ifx \showISBNx    \undefined \def \showISBNx     #1{\unskip}     \fi
\ifx \showISBNxiii \undefined \def \showISBNxiii  #1{\unskip}     \fi
\ifx \showISSN     \undefined \def \showISSN      #1{\unskip}     \fi
\ifx \showLCCN     \undefined \def \showLCCN      #1{\unskip}     \fi
\ifx \shownote     \undefined \def \shownote      #1{#1}          \fi
\ifx \showarticletitle \undefined \def \showarticletitle #1{#1}   \fi
\ifx \showURL      \undefined \def \showURL       {\relax}        \fi
% invisible to TeX
\providecommand\bibfield[2]{#2}
\providecommand\bibinfo[2]{#2}
\providecommand\natexlab[1]{#1}
\providecommand\showeprint[2][]{arXiv:#2}

\bibitem[Adamczyk and Bailey(2004)]%
        {adamczyk2004if}
\bibfield{author}{\bibinfo{person}{Piotr~D Adamczyk} {and}
  \bibinfo{person}{Brian~P Bailey}.} \bibinfo{year}{2004}\natexlab{}.
\newblock \showarticletitle{If not now, when? The effects of interruption at
  different moments within task execution}. In
  \bibinfo{booktitle}{\emph{Proceedings of the SIGCHI conference on Human
  factors in computing systems}}. \bibinfo{pages}{271--278}.
\newblock
\urldef\tempurl%
\url{https://doi.org/10.1145/985692.985727}
\showDOI{\tempurl}


\bibitem[Addas and Pinsonneault(2018)]%
        {addas2018theorizing}
\bibfield{author}{\bibinfo{person}{Shamel Addas} {and} \bibinfo{person}{Alain
  Pinsonneault}.} \bibinfo{year}{2018}\natexlab{}.
\newblock \showarticletitle{Theorizing the {Multilevel} {Effects} of
  {Interruptions} and the {Role} of {Communication} {Technology}}.
\newblock \bibinfo{journal}{\emph{Journal of the Association for Information
  Systems}} \bibinfo{volume}{19}, \bibinfo{number}{11} (\bibinfo{date}{Nov.}
  \bibinfo{year}{2018}).
\newblock
\showISSN{1536-9323}
\urldef\tempurl%
\url{https://aisel.aisnet.org/jais/vol19/iss11/2}
\showURL{%
\tempurl}


\bibitem[Allen et~al\mbox{.}(2014)]%
        {allen2014multitasking}
\bibfield{author}{\bibinfo{person}{Philip~A Allen}, \bibinfo{person}{Mei-Ching
  Lien}, \bibinfo{person}{Eric Ruthruff}, {and} \bibinfo{person}{Andreas
  Voss}.} \bibinfo{year}{2014}\natexlab{}.
\newblock \showarticletitle{Multitasking and aging: do older adults benefit
  from performing a highly practiced task?}
\newblock \bibinfo{journal}{\emph{Experimental aging research}}
  \bibinfo{volume}{40}, \bibinfo{number}{3} (\bibinfo{year}{2014}),
  \bibinfo{pages}{280--307}.
\newblock
\urldef\tempurl%
\url{https://doi.org/10.1080/0361073X.2014.896663}
\showDOI{\tempurl}


\bibitem[Altmann and Schunn(2019)]%
        {altmann2019integrating}
\bibfield{author}{\bibinfo{person}{Erik~M Altmann} {and}
  \bibinfo{person}{Christian~D Schunn}.} \bibinfo{year}{2019}\natexlab{}.
\newblock \showarticletitle{Integrating decay and interference: A new look at
  an old interaction}. In \bibinfo{booktitle}{\emph{Proceedings of the
  Twenty-Fourth Annual Conference of the Cognitive Science Society}}.
  Routledge, \bibinfo{pages}{65--70}.
\newblock
\urldef\tempurl%
\url{https://doi.org/10.4324/9781315782379-49}
\showDOI{\tempurl}


\bibitem[Altmann and Trafton(2002)]%
        {altmann2002memory}
\bibfield{author}{\bibinfo{person}{Erik~M Altmann} {and}
  \bibinfo{person}{J~Gregory Trafton}.} \bibinfo{year}{2002}\natexlab{}.
\newblock \showarticletitle{Memory for goals: An activation-based model}.
\newblock \bibinfo{journal}{\emph{Cognitive science}} \bibinfo{volume}{26},
  \bibinfo{number}{1} (\bibinfo{year}{2002}), \bibinfo{pages}{39--83}.
\newblock
\urldef\tempurl%
\url{https://doi.org/10.1207/s15516709cog2601_2}
\showDOI{\tempurl}


\bibitem[Altmann et~al\mbox{.}(2014)]%
        {altmann2014momentary}
\bibfield{author}{\bibinfo{person}{Erik~M Altmann}, \bibinfo{person}{J~Gregory
  Trafton}, {and} \bibinfo{person}{David~Z Hambrick}.}
  \bibinfo{year}{2014}\natexlab{}.
\newblock \showarticletitle{Momentary interruptions can derail the train of
  thought.}
\newblock \bibinfo{journal}{\emph{Journal of Experimental Psychology: General}}
  \bibinfo{volume}{143}, \bibinfo{number}{1} (\bibinfo{year}{2014}),
  \bibinfo{pages}{215}.
\newblock
\urldef\tempurl%
\url{https://doi.org/10.1037/a0030986}
\showDOI{\tempurl}


\bibitem[Ball and Aschenbrenner(2018)]%
        {ball2018importance}
\bibfield{author}{\bibinfo{person}{B~Hunter Ball} {and}
  \bibinfo{person}{Andrew~J Aschenbrenner}.} \bibinfo{year}{2018}\natexlab{}.
\newblock \showarticletitle{The importance of age-related differences in
  prospective memory: Evidence from diffusion model analyses}.
\newblock \bibinfo{journal}{\emph{Psychonomic Bulletin \& Review}}
  \bibinfo{volume}{25}, \bibinfo{number}{3} (\bibinfo{year}{2018}),
  \bibinfo{pages}{1114--1122}.
\newblock
\urldef\tempurl%
\url{https://doi.org/10.3758/s13423-017-1318-4}
\showDOI{\tempurl}


\bibitem[Barr et~al\mbox{.}(2013)]%
        {barr2013random}
\bibfield{author}{\bibinfo{person}{Dale~J. Barr}, \bibinfo{person}{Roger Levy},
  \bibinfo{person}{Christoph Scheepers}, {and} \bibinfo{person}{Harry~J.
  Tily}.} \bibinfo{year}{2013}\natexlab{}.
\newblock \showarticletitle{Random effects structure for confirmatory
  hypothesis testing: Keep it maximal}.
\newblock \bibinfo{journal}{\emph{Journal of Memory and Language}}
  \bibinfo{volume}{68}, \bibinfo{number}{3} (\bibinfo{year}{2013}),
  \bibinfo{pages}{255--278}.
\newblock
\showISSN{0749-596X}
\urldef\tempurl%
\url{https://doi.org/10.1016/j.jml.2012.11.001}
\showDOI{\tempurl}


\bibitem[Boag et~al\mbox{.}(2019)]%
        {boag2019strategic}
\bibfield{author}{\bibinfo{person}{Russell~J Boag}, \bibinfo{person}{Luke
  Strickland}, \bibinfo{person}{Shayne Loft}, {and} \bibinfo{person}{Andrew
  Heathcote}.} \bibinfo{year}{2019}\natexlab{}.
\newblock \showarticletitle{Strategic attention and decision control support
  prospective memory in a complex dual-task environment}.
\newblock \bibinfo{journal}{\emph{Cognition}}  \bibinfo{volume}{191}
  (\bibinfo{year}{2019}), \bibinfo{pages}{103974}.
\newblock
\urldef\tempurl%
\url{https://doi.org/10.1016/j.cognition.2019.05.011}
\showDOI{\tempurl}


\bibitem[Bobadilla et~al\mbox{.}(2013)]%
        {bobadilla2013recommender}
\bibfield{author}{\bibinfo{person}{Jes{\'u}s Bobadilla},
  \bibinfo{person}{Fernando Ortega}, \bibinfo{person}{Antonio Hernando}, {and}
  \bibinfo{person}{Abraham Guti{\'e}rrez}.} \bibinfo{year}{2013}\natexlab{}.
\newblock \showarticletitle{Recommender systems survey}.
\newblock \bibinfo{journal}{\emph{Knowledge-based systems}}
  \bibinfo{volume}{46} (\bibinfo{year}{2013}), \bibinfo{pages}{109--132}.
\newblock
\urldef\tempurl%
\url{https://doi.org/ttps://doi.org/10.1016/j.knosys.2013.03.012}
\showDOI{\tempurl}


\bibitem[Boywitt and Rummel(2012)]%
        {boywitt2012diffusion}
\bibfield{author}{\bibinfo{person}{C~Dennis Boywitt} {and} \bibinfo{person}{Jan
  Rummel}.} \bibinfo{year}{2012}\natexlab{}.
\newblock \showarticletitle{A diffusion model analysis of task interference
  effects in prospective memory}.
\newblock \bibinfo{journal}{\emph{Memory \& Cognition}} \bibinfo{volume}{40},
  \bibinfo{number}{1} (\bibinfo{year}{2012}), \bibinfo{pages}{70--82}.
\newblock
\urldef\tempurl%
\url{https://doi.org/10.3758/s13421-011-0128-6}
\showDOI{\tempurl}


\bibitem[Brewer(2011)]%
        {brewer2011analyzing}
\bibfield{author}{\bibinfo{person}{Gene~A Brewer}.}
  \bibinfo{year}{2011}\natexlab{}.
\newblock \showarticletitle{Analyzing response time distributions:
  Methodological and theoretical suggestions for prospective memory
  researchers.}
\newblock \bibinfo{journal}{\emph{Zeitschrift f{\"u}r Psychologie/Journal of
  Psychology}} \bibinfo{volume}{219}, \bibinfo{number}{2}
  (\bibinfo{year}{2011}), \bibinfo{pages}{117}.
\newblock
\urldef\tempurl%
\url{https://doi.org/10.1027/2151-2604/a000056}
\showDOI{\tempurl}


\bibitem[Brewer et~al\mbox{.}(2017)]%
        {brewer2017how}
\bibfield{author}{\bibinfo{person}{R.~N. Brewer}, \bibinfo{person}{M.~R.
  Morris}, {and} \bibinfo{person}{S.~E. Lindley}.}
  \bibinfo{year}{2017}\natexlab{}.
\newblock \showarticletitle{How to Remember What to Remember: Exploring
  Possibilities for Digital Reminder Systems}.
\newblock  \bibinfo{volume}{1}, \bibinfo{number}{3}, Article
  \bibinfo{articleno}{38} (\bibinfo{date}{sep} \bibinfo{year}{2017}),
  \bibinfo{numpages}{20}~pages.
\newblock
\urldef\tempurl%
\url{https://doi.org/10.1145/3130903}
\showDOI{\tempurl}


\bibitem[Brumby et~al\mbox{.}(2019)]%
        {brumby2019interruptions}
\bibfield{author}{\bibinfo{person}{Duncan~P Brumby},
  \bibinfo{person}{Christian~P Janssen}, {and} \bibinfo{person}{Gloria Mark}.}
  \bibinfo{year}{2019}\natexlab{}.
\newblock \showarticletitle{How do interruptions affect productivity?}
\newblock In \bibinfo{booktitle}{\emph{Rethinking productivity in software
  engineering}}. \bibinfo{publisher}{Springer}, \bibinfo{pages}{85--107}.
\newblock
\urldef\tempurl%
\url{https://doi.org/10.1016/j.ijhcs.2014.11.003}
\showDOI{\tempurl}


\bibitem[Brysbaert et~al\mbox{.}(2011)]%
        {brysbaert2011word}
\bibfield{author}{\bibinfo{person}{Marc Brysbaert}, \bibinfo{person}{Matthias
  Buchmeier}, \bibinfo{person}{Markus Conrad}, \bibinfo{person}{Arthur~M
  Jacobs}, \bibinfo{person}{Jens B{\"o}lte}, {and} \bibinfo{person}{Andrea
  B{\"o}hl}.} \bibinfo{year}{2011}\natexlab{}.
\newblock \showarticletitle{The word frequency effect: a review of recent
  developments and implications for the choice of frequency estimates in
  German.}
\newblock \bibinfo{journal}{\emph{Experimental psychology}}
  \bibinfo{volume}{58}, \bibinfo{number}{5} (\bibinfo{year}{2011}),
  \bibinfo{pages}{412}.
\newblock
\urldef\tempurl%
\url{https://doi.org/10.1027/1618-3169/a000123}
\showDOI{\tempurl}


\bibitem[Busch et~al\mbox{.}(2022)]%
        {busch2022german}
\bibfield{author}{\bibinfo{person}{Johannes~L Busch}, \bibinfo{person}{Femke~S
  Haeussler}, \bibinfo{person}{Frank Domahs}, \bibinfo{person}{Lars
  Timmermann}, \bibinfo{person}{Immo Weber}, {and} \bibinfo{person}{Carina~R
  Oehrn}.} \bibinfo{year}{2022}\natexlab{}.
\newblock \showarticletitle{German normative data with naming latencies for 283
  action pictures and 600 action verbs}.
\newblock \bibinfo{journal}{\emph{Behavior research methods}}
  \bibinfo{volume}{54}, \bibinfo{number}{2} (\bibinfo{year}{2022}),
  \bibinfo{pages}{649--662}.
\newblock
\urldef\tempurl%
\url{https://doi.org/10.3758/s13428-021-01647-w}
\showDOI{\tempurl}


\bibitem[Chen et~al\mbox{.}(2019)]%
        {chen2019study}
\bibfield{author}{\bibinfo{person}{Zhuang Chen}, \bibinfo{person}{Qian He},
  \bibinfo{person}{Zhifei Mao}, \bibinfo{person}{Hwei-Ming Chung}, {and}
  \bibinfo{person}{Sabita Maharjan}.} \bibinfo{year}{2019}\natexlab{}.
\newblock \showarticletitle{A study on the characteristics of douyin short
  videos and implications for edge caching}. In
  \bibinfo{booktitle}{\emph{Proceedings of the ACM Turing Celebration
  Conference-China}}. \bibinfo{pages}{1--6}.
\newblock
\urldef\tempurl%
\url{https://doi.org/10.1145/3321408.3323082}
\showDOI{\tempurl}


\bibitem[Chong and Siino(2006)]%
        {chong2006interruptions}
\bibfield{author}{\bibinfo{person}{Jan Chong} {and} \bibinfo{person}{Rosanne
  Siino}.} \bibinfo{year}{2006}\natexlab{}.
\newblock \showarticletitle{Interruptions on software teams: a comparison of
  paired and solo programmers}. In \bibinfo{booktitle}{\emph{Proceedings of the
  2006 20th anniversary conference on Computer supported cooperative work}}.
  \bibinfo{pages}{29--38}.
\newblock
\urldef\tempurl%
\url{https://doi.org/10.1145/1180875.1180882}
\showDOI{\tempurl}


\bibitem[Clapp and Gazzaley(2012)]%
        {clapp2012distinct}
\bibfield{author}{\bibinfo{person}{Wesley~C Clapp} {and} \bibinfo{person}{Adam
  Gazzaley}.} \bibinfo{year}{2012}\natexlab{}.
\newblock \showarticletitle{Distinct mechanisms for the impact of distraction
  and interruption on working memory in aging}.
\newblock \bibinfo{journal}{\emph{Neurobiology of aging}} \bibinfo{volume}{33},
  \bibinfo{number}{1} (\bibinfo{year}{2012}), \bibinfo{pages}{134--148}.
\newblock
\urldef\tempurl%
\url{https://doi.org/10.1016/j.neurobiolaging.2010.01.012}
\showDOI{\tempurl}


\bibitem[Cohen and Hicks(2017)]%
        {cohen2017characterization}
\bibfield{author}{\bibinfo{person}{Anna-Lisa Cohen} {and}
  \bibinfo{person}{Jason~L. Hicks}.} \bibinfo{year}{2017}\natexlab{}.
\newblock \bibinfo{booktitle}{\emph{Characterization of Prospective Memory and
  Associated Processes}}.
\newblock \bibinfo{publisher}{Springer International Publishing},
  \bibinfo{address}{Cham}, \bibinfo{pages}{41–60}.
\newblock
\showISBNx{978-3-319-68990-6}
\urldef\tempurl%
\url{https://doi.org/10.1007/978-3-319-68990-6_3}
\showDOI{\tempurl}


\bibitem[Colder~Carras et~al\mbox{.}(2018)]%
        {colder_carras_connection_2018}
\bibfield{author}{\bibinfo{person}{Michelle Colder~Carras},
  \bibinfo{person}{Anna Kalbarczyk}, \bibinfo{person}{Kurrie Wells},
  \bibinfo{person}{Jaime Banks}, \bibinfo{person}{Rachel Kowert},
  \bibinfo{person}{Colleen Gillespie}, {and} \bibinfo{person}{Carl Latkin}.}
  \bibinfo{year}{2018}\natexlab{}.
\newblock \showarticletitle{Connection, meaning, and distraction: {A}
  qualitative study of video game play and mental health recovery in veterans
  treated for mental and/or behavioral health problems}.
\newblock \bibinfo{journal}{\emph{Social Science \& Medicine}}
  \bibinfo{volume}{216} (\bibinfo{date}{Nov.} \bibinfo{year}{2018}),
  \bibinfo{pages}{124--132}.
\newblock
\showISSN{0277-9536}
\urldef\tempurl%
\url{https://doi.org/10.1016/j.socscimed.2018.08.044}
\showDOI{\tempurl}


\bibitem[Cona et~al\mbox{.}(2020)]%
        {cona_theta_2020}
\bibfield{author}{\bibinfo{person}{Giorgia Cona}, \bibinfo{person}{Francesco
  Chiossi}, \bibinfo{person}{Silvia Di~Tomasso}, \bibinfo{person}{Giovanni
  Pellegrino}, \bibinfo{person}{Francesco Piccione}, \bibinfo{person}{Patrizia
  Bisiacchi}, {and} \bibinfo{person}{Giorgio Arcara}.}
  \bibinfo{year}{2020}\natexlab{}.
\newblock \showarticletitle{Theta and alpha oscillations as signatures of
  internal and external attention to delayed intentions: {A}
  magnetoencephalography ({MEG}) study}.
\newblock \bibinfo{journal}{\emph{NeuroImage}}  \bibinfo{volume}{205}
  (\bibinfo{date}{Jan.} \bibinfo{year}{2020}), \bibinfo{pages}{116295}.
\newblock
\showISSN{1053-8119}
\urldef\tempurl%
\url{https://doi.org/10.1016/j.neuroimage.2019.116295}
\showDOI{\tempurl}


\bibitem[Cona et~al\mbox{.}(2015)]%
        {cona2015neural}
\bibfield{author}{\bibinfo{person}{Giorgia Cona}, \bibinfo{person}{Cristina
  Scarpazza}, \bibinfo{person}{Giuseppe Sartori}, \bibinfo{person}{Morris
  Moscovitch}, {and} \bibinfo{person}{Patrizia~Silvia Bisiacchi}.}
  \bibinfo{year}{2015}\natexlab{}.
\newblock \showarticletitle{Neural bases of prospective memory: a meta-analysis
  and the “Attention to Delayed Intention”(AtoDI) model}.
\newblock \bibinfo{journal}{\emph{Neuroscience \& Biobehavioral Reviews}}
  \bibinfo{volume}{52} (\bibinfo{year}{2015}), \bibinfo{pages}{21--37}.
\newblock
\urldef\tempurl%
\url{https://doi.org/10.1016/j.neubiorev.2015.02.007}
\showDOI{\tempurl}


\bibitem[Cook et~al\mbox{.}(2014)]%
        {cook2014role}
\bibfield{author}{\bibinfo{person}{Gabriel~I Cook}, \bibinfo{person}{J~Thadeus
  Meeks}, \bibinfo{person}{Arlo Clark-Foos}, \bibinfo{person}{Paul~S Merritt},
  {and} \bibinfo{person}{Richard~L Marsh}.} \bibinfo{year}{2014}\natexlab{}.
\newblock \showarticletitle{The role of interruptions and contextual
  associations in delayed-execute prospective memory}.
\newblock \bibinfo{journal}{\emph{Applied Cognitive Psychology}}
  \bibinfo{volume}{28}, \bibinfo{number}{1} (\bibinfo{year}{2014}),
  \bibinfo{pages}{91--103}.
\newblock
\urldef\tempurl%
\url{https://doi.org/10.1002/acp.2960}
\showDOI{\tempurl}


\bibitem[Craik et~al\mbox{.}(1996)]%
        {craik_effects_1996}
\bibfield{author}{\bibinfo{person}{Fergus I.~M. Craik},
  \bibinfo{person}{Richard Govoni}, \bibinfo{person}{Moshe Naveh-Benjamin},
  {and} \bibinfo{person}{Nicole~D. Anderson}.} \bibinfo{year}{1996}\natexlab{}.
\newblock \showarticletitle{The effects of divided attention on encoding and
  retrieval processes in human memory.}
\newblock \bibinfo{journal}{\emph{Journal of Experimental Psychology: General}}
  \bibinfo{volume}{125}, \bibinfo{number}{2} (\bibinfo{year}{1996}),
  \bibinfo{pages}{159--180}.
\newblock
\showISSN{1939-2222, 0096-3445}
\urldef\tempurl%
\url{https://doi.org/10.1037/0096-3445.125.2.159}
\showDOI{\tempurl}


\bibitem[Davidson et~al\mbox{.}(2010)]%
        {davidson2010youtube}
\bibfield{author}{\bibinfo{person}{James Davidson}, \bibinfo{person}{Benjamin
  Liebald}, \bibinfo{person}{Junning Liu}, \bibinfo{person}{Palash Nandy},
  \bibinfo{person}{Taylor Van~Vleet}, \bibinfo{person}{Ullas Gargi},
  \bibinfo{person}{Sujoy Gupta}, \bibinfo{person}{Yu He}, \bibinfo{person}{Mike
  Lambert}, \bibinfo{person}{Blake Livingston}, {et~al\mbox{.}}}
  \bibinfo{year}{2010}\natexlab{}.
\newblock \showarticletitle{The YouTube video recommendation system}. In
  \bibinfo{booktitle}{\emph{Proceedings of the fourth ACM conference on
  Recommender systems}}. \bibinfo{pages}{293--296}.
\newblock
\urldef\tempurl%
\url{https://doi.org/10.1145/1864708.1864770}
\showDOI{\tempurl}


\bibitem[Dismukes(2012)]%
        {dismukes2012prospective}
\bibfield{author}{\bibinfo{person}{R~Key Dismukes}.}
  \bibinfo{year}{2012}\natexlab{}.
\newblock \showarticletitle{Prospective memory in workplace and everyday
  situations}.
\newblock \bibinfo{journal}{\emph{Current Directions in Psychological Science}}
  \bibinfo{volume}{21}, \bibinfo{number}{4} (\bibinfo{year}{2012}),
  \bibinfo{pages}{215--220}.
\newblock
\urldef\tempurl%
\url{https://doi.org/10.1177/0963721412447621}
\showDOI{\tempurl}


\bibitem[Dodhia and Dismukes(2009)]%
        {dodhia2009interruptions}
\bibfield{author}{\bibinfo{person}{Rahul~M Dodhia} {and}
  \bibinfo{person}{Robert~K Dismukes}.} \bibinfo{year}{2009}\natexlab{}.
\newblock \showarticletitle{Interruptions create prospective memory tasks}.
\newblock \bibinfo{journal}{\emph{Applied Cognitive Psychology: The Official
  Journal of the Society for Applied Research in Memory and Cognition}}
  \bibinfo{volume}{23}, \bibinfo{number}{1} (\bibinfo{year}{2009}),
  \bibinfo{pages}{73--89}.
\newblock
\urldef\tempurl%
\url{https://doi.org/10.1002/acp.1441}
\showDOI{\tempurl}


\bibitem[Doost and Zhang(2022)]%
        {zahmat2022mental}
\bibfield{author}{\bibinfo{person}{Elmira~Zahmat Doost} {and}
  \bibinfo{person}{Wei Zhang}.} \bibinfo{year}{2022}\natexlab{}.
\newblock \showarticletitle{Mental workload variations during different
  cognitive office tasks with social media interruptions}.
\newblock \bibinfo{journal}{\emph{Ergonomics}} \bibinfo{volume}{0},
  \bibinfo{number}{0} (\bibinfo{year}{2022}), \bibinfo{pages}{1--17}.
\newblock
\urldef\tempurl%
\url{https://doi.org/10.1080/00140139.2022.2104381}
\showDOI{\tempurl}


\bibitem[Duinkharjav et~al\mbox{.}(2022)]%
        {duinkharjav2022image}
\bibfield{author}{\bibinfo{person}{Budmonde Duinkharjav},
  \bibinfo{person}{Praneeth Chakravarthula}, \bibinfo{person}{Rachel Brown},
  \bibinfo{person}{Anjul Patney}, {and} \bibinfo{person}{Qi Sun}.}
  \bibinfo{year}{2022}\natexlab{}.
\newblock \showarticletitle{Image Features Influence Reaction Time: A Learned
  Probabilistic Perceptual Model for Saccade Latency}.
\newblock \bibinfo{journal}{\emph{ACM Trans. Graph.}} \bibinfo{volume}{41},
  \bibinfo{number}{4}, Article \bibinfo{articleno}{144} (\bibinfo{date}{jul}
  \bibinfo{year}{2022}), \bibinfo{numpages}{15}~pages.
\newblock
\showISSN{0730-0301}
\urldef\tempurl%
\url{https://doi.org/10.1145/3528223.3530055}
\showDOI{\tempurl}


\bibitem[Einstein and McDaniel(2005)]%
        {einstein2005prospective}
\bibfield{author}{\bibinfo{person}{Gilles~O Einstein} {and}
  \bibinfo{person}{Mark~A McDaniel}.} \bibinfo{year}{2005}\natexlab{}.
\newblock \showarticletitle{Prospective memory: Multiple retrieval processes}.
\newblock \bibinfo{journal}{\emph{Current Directions in Psychological Science}}
  \bibinfo{volume}{14}, \bibinfo{number}{6} (\bibinfo{year}{2005}),
  \bibinfo{pages}{286--290}.
\newblock
\urldef\tempurl%
\url{https://doi.org/10.1111/j.0963-7214.2005.00382.x}
\showDOI{\tempurl}


\bibitem[Einstein et~al\mbox{.}(2005a)]%
        {einstein2005multiple}
\bibfield{author}{\bibinfo{person}{Gilles~O Einstein}, \bibinfo{person}{Mark~A
  McDaniel}, \bibinfo{person}{Ruthann Thomas}, \bibinfo{person}{Sara Mayfield},
  \bibinfo{person}{Hilary Shank}, \bibinfo{person}{Nova Morrisette}, {and}
  \bibinfo{person}{Jennifer Breneiser}.} \bibinfo{year}{2005}\natexlab{a}.
\newblock \showarticletitle{Multiple processes in prospective memory retrieval:
  factors determining monitoring versus spontaneous retrieval.}
\newblock \bibinfo{journal}{\emph{Journal of Experimental Psychology: General}}
  \bibinfo{volume}{134}, \bibinfo{number}{3} (\bibinfo{year}{2005}),
  \bibinfo{pages}{327}.
\newblock
\urldef\tempurl%
\url{https://doi.org/10.1037/0096-3445.134.3.327}
\showDOI{\tempurl}


\bibitem[Einstein et~al\mbox{.}(2005b)]%
        {einstein_multiple_2005}
\bibfield{author}{\bibinfo{person}{Gilles~O. Einstein},
  \bibinfo{person}{Mark~A. McDaniel}, \bibinfo{person}{Ruthann Thomas},
  \bibinfo{person}{Sara Mayfield}, \bibinfo{person}{Hilary Shank},
  \bibinfo{person}{Nova Morrisette}, {and} \bibinfo{person}{Jennifer
  Breneiser}.} \bibinfo{year}{2005}\natexlab{b}.
\newblock \showarticletitle{Multiple {Processes} in {Prospective} {Memory}
  {Retrieval}: {Factors} {Determining} {Monitoring} {Versus} {Spontaneous}
  {Retrieval}}.
\newblock \bibinfo{journal}{\emph{Journal of Experimental Psychology: General}}
  \bibinfo{volume}{134}, \bibinfo{number}{3} (\bibinfo{year}{2005}),
  \bibinfo{pages}{327--342}.
\newblock
\showISSN{1939-2222}
\urldef\tempurl%
\url{https://doi.org/10.1037/0096-3445.134.3.327}
\showDOI{\tempurl}


\bibitem[Einstein et~al\mbox{.}(1997)]%
        {einstein1997aging}
\bibfield{author}{\bibinfo{person}{Gilles~O Einstein},
  \bibinfo{person}{Rebekah~E Smith}, \bibinfo{person}{Mark~A McDaniel}, {and}
  \bibinfo{person}{Pat Shaw}.} \bibinfo{year}{1997}\natexlab{}.
\newblock \showarticletitle{Aging and prospective memory: the influence of
  increased task demands at encoding and retrieval.}
\newblock \bibinfo{journal}{\emph{Psychology and aging}} \bibinfo{volume}{12},
  \bibinfo{number}{3} (\bibinfo{year}{1997}), \bibinfo{pages}{479}.
\newblock
\urldef\tempurl%
\url{https://doi.org/10.1037/0882-7974.12.3.479}
\showDOI{\tempurl}


\bibitem[Fox et~al\mbox{.}(2009)]%
        {fox2009distractions}
\bibfield{author}{\bibinfo{person}{Annie~Beth Fox}, \bibinfo{person}{Jonathan
  Rosen}, {and} \bibinfo{person}{Mary Crawford}.}
  \bibinfo{year}{2009}\natexlab{}.
\newblock \showarticletitle{Distractions, distractions: does instant messaging
  affect college students' performance on a concurrent reading comprehension
  task?}
\newblock \bibinfo{journal}{\emph{CyberPsychology \& Behavior}}
  \bibinfo{volume}{12}, \bibinfo{number}{1} (\bibinfo{year}{2009}),
  \bibinfo{pages}{51--53}.
\newblock
\urldef\tempurl%
\url{https://doi.org/10.1089/cpb.2008.0107}
\showDOI{\tempurl}


\bibitem[Frein et~al\mbox{.}(2013)]%
        {frein2013comes}
\bibfield{author}{\bibinfo{person}{Scott~T Frein}, \bibinfo{person}{Samantha~L
  Jones}, {and} \bibinfo{person}{Jennifer~E Gerow}.}
  \bibinfo{year}{2013}\natexlab{}.
\newblock \showarticletitle{When it comes to Facebook there may be more to bad
  memory than just multitasking}.
\newblock \bibinfo{journal}{\emph{Computers in Human Behavior}}
  \bibinfo{volume}{29}, \bibinfo{number}{6} (\bibinfo{year}{2013}),
  \bibinfo{pages}{2179--2182}.
\newblock
\urldef\tempurl%
\url{https://doi.org/10.1016/j.chb.2013.04.031}
\showDOI{\tempurl}


\bibitem[Gordon et~al\mbox{.}(2011)]%
        {gordon2011structural}
\bibfield{author}{\bibinfo{person}{Brian~A Gordon}, \bibinfo{person}{Jill~T
  Shelton}, \bibinfo{person}{Julie~M Bugg}, \bibinfo{person}{Mark~A McDaniel},
  {and} \bibinfo{person}{Denise Head}.} \bibinfo{year}{2011}\natexlab{}.
\newblock \showarticletitle{Structural correlates of prospective memory}.
\newblock \bibinfo{journal}{\emph{Neuropsychologia}} \bibinfo{volume}{49},
  \bibinfo{number}{14} (\bibinfo{year}{2011}), \bibinfo{pages}{3795--3800}.
\newblock
\urldef\tempurl%
\url{https://doi.org/10.1016/j.neuropsychologia.2011.09.035}
\showDOI{\tempurl}


\bibitem[Gray et~al\mbox{.}(2018)]%
        {gray_dark_2018}
\bibfield{author}{\bibinfo{person}{Colin~M. Gray}, \bibinfo{person}{Yubo Kou},
  \bibinfo{person}{Bryan Battles}, \bibinfo{person}{Joseph Hoggatt}, {and}
  \bibinfo{person}{Austin~L. Toombs}.} \bibinfo{year}{2018}\natexlab{}.
\newblock \showarticletitle{The {Dark} ({Patterns}) {Side} of {UX} {Design}}.
  In \bibinfo{booktitle}{\emph{Proceedings of the 2018 {CHI} {Conference} on
  {Human} {Factors} in {Computing} {Systems}}} \emph{(\bibinfo{series}{{CHI}
  '18})}. \bibinfo{publisher}{Association for Computing Machinery},
  \bibinfo{address}{New York, NY, USA}, \bibinfo{pages}{1--14}.
\newblock
\showISBNx{978-1-4503-5620-6}
\urldef\tempurl%
\url{https://doi.org/10.1145/3173574.3174108}
\showDOI{\tempurl}


\bibitem[Groot et~al\mbox{.}(2002)]%
        {groot_prospective_2002}
\bibfield{author}{\bibinfo{person}{Yvonne C.~T. Groot},
  \bibinfo{person}{Barbara~A. Wilson}, \bibinfo{person}{Jonathan Evans}, {and}
  \bibinfo{person}{Peter Watson}.} \bibinfo{year}{2002}\natexlab{}.
\newblock \showarticletitle{Prospective memory functioning in people with and
  without brain injury}.
\newblock \bibinfo{journal}{\emph{Journal of the International
  Neuropsychological Society}} \bibinfo{volume}{8}, \bibinfo{number}{5}
  (\bibinfo{date}{July} \bibinfo{year}{2002}), \bibinfo{pages}{645--654}.
\newblock
\showISSN{1355-6177, 1469-7661}
\urldef\tempurl%
\url{https://doi.org/10.1017/S1355617702801321}
\showDOI{\tempurl}


\bibitem[Gupta et~al\mbox{.}(2013)]%
        {gupta2013should}
\bibfield{author}{\bibinfo{person}{Ashish Gupta}, \bibinfo{person}{Han Li},
  {and} \bibinfo{person}{Ramesh Sharda}.} \bibinfo{year}{2013}\natexlab{}.
\newblock \showarticletitle{Should I send this message? Understanding the
  impact of interruptions, social hierarchy and perceived task complexity on
  user performance and perceived workload}.
\newblock \bibinfo{journal}{\emph{Decision Support Systems}}
  \bibinfo{volume}{55}, \bibinfo{number}{1} (\bibinfo{year}{2013}),
  \bibinfo{pages}{135--145}.
\newblock
\urldef\tempurl%
\url{https://doi.org/10.1016/j.dss.2012.12.035}
\showDOI{\tempurl}


\bibitem[Guynn(2003)]%
        {guynn2003two}
\bibfield{author}{\bibinfo{person}{Melissa~J Guynn}.}
  \bibinfo{year}{2003}\natexlab{}.
\newblock \showarticletitle{A two-process model of strategic monitoring in
  event-based prospective memory: Activation/retrieval mode and checking}.
\newblock \bibinfo{journal}{\emph{International Journal of Psychology}}
  \bibinfo{volume}{38}, \bibinfo{number}{4} (\bibinfo{year}{2003}),
  \bibinfo{pages}{245--256}.
\newblock
\urldef\tempurl%
\url{https://doi.org/10.1080/00207590344000178}
\showDOI{\tempurl}


\bibitem[Guynn et~al\mbox{.}(1998)]%
        {guynn_prospective_1998}
\bibfield{author}{\bibinfo{person}{Melissa~J. Guynn}, \bibinfo{person}{Mark~A.
  Mcdaniel}, {and} \bibinfo{person}{Gilles~O. Einstein}.}
  \bibinfo{year}{1998}\natexlab{}.
\newblock \showarticletitle{Prospective memory: {When} reminders fail}.
\newblock \bibinfo{journal}{\emph{Memory \& Cognition}} \bibinfo{volume}{26},
  \bibinfo{number}{2} (\bibinfo{date}{March} \bibinfo{year}{1998}),
  \bibinfo{pages}{287--298}.
\newblock
\showISSN{1532-5946}
\urldef\tempurl%
\url{https://doi.org/10.3758/BF03201140}
\showDOI{\tempurl}


\bibitem[Hefner and Vorderer(2016)]%
        {hefner_digital_2016}
\bibfield{author}{\bibinfo{person}{Dorothee Hefner} {and}
  \bibinfo{person}{Peter Vorderer}.} \bibinfo{year}{2016}\natexlab{}.
\newblock \showarticletitle{Digital {Stress}: {Permanent} {Connectedness} and
  {Multitasking}}.
\newblock In \bibinfo{booktitle}{\emph{The {Routledge} {Handbook} of {Media}
  {Use} and {Well}-{Being}: {International} {Perspectives} on {Theory} and
  {Research} on {Positive} {Media} {Effects}.}} \bibinfo{publisher}{Taylor \&
  Francis}, \bibinfo{pages}{484}.
\newblock
\showISBNx{978-0-367-73699-6}
\urldef\tempurl%
\url{https://doi.org/10.4324/9781315714752}
\showDOI{\tempurl}


\bibitem[Hiniker et~al\mbox{.}(2016)]%
        {hiniker_mytime_2016}
\bibfield{author}{\bibinfo{person}{Alexis Hiniker},
  \bibinfo{person}{Sungsoo~(Ray) Hong}, \bibinfo{person}{Tadayoshi Kohno},
  {and} \bibinfo{person}{Julie~A. Kientz}.} \bibinfo{year}{2016}\natexlab{}.
\newblock \showarticletitle{{MyTime}: {Designing} and {Evaluating} an
  {Intervention} for {Smartphone} {Non}-{Use}}. In
  \bibinfo{booktitle}{\emph{Proceedings of the 2016 {CHI} {Conference} on
  {Human} {Factors} in {Computing} {Systems}}} \emph{(\bibinfo{series}{{CHI}
  '16})}. \bibinfo{publisher}{Association for Computing Machinery},
  \bibinfo{address}{New York, NY, USA}, \bibinfo{pages}{4746--4757}.
\newblock
\showISBNx{978-1-4503-3362-7}
\urldef\tempurl%
\url{https://doi.org/10.1145/2858036.2858403}
\showDOI{\tempurl}


\bibitem[Hong and Leber(2016)]%
        {hong2016attentional}
\bibfield{author}{\bibinfo{person}{Yoolim Hong} {and} \bibinfo{person}{Andrew
  Leber}.} \bibinfo{year}{2016}\natexlab{}.
\newblock \showarticletitle{Attentional disengagement suppresses visual
  long-term memory}.
\newblock \bibinfo{journal}{\emph{Journal of Vision}} \bibinfo{volume}{16},
  \bibinfo{number}{12} (\bibinfo{year}{2016}), \bibinfo{pages}{1017--1017}.
\newblock
\urldef\tempurl%
\url{https://doi.org/10.1167/16.12.1017}
\showDOI{\tempurl}


\bibitem[Horn et~al\mbox{.}(2011)]%
        {horn2011can}
\bibfield{author}{\bibinfo{person}{Sebastian~S Horn}, \bibinfo{person}{Ute~J
  Bayen}, {and} \bibinfo{person}{Rebekah~E Smith}.}
  \bibinfo{year}{2011}\natexlab{}.
\newblock \showarticletitle{What can the diffusion model tell us about
  prospective memory?}
\newblock \bibinfo{journal}{\emph{Canadian Journal of Experimental
  Psychology/Revue canadienne de psychologie exp{\'e}rimentale}}
  \bibinfo{volume}{65}, \bibinfo{number}{1} (\bibinfo{year}{2011}),
  \bibinfo{pages}{69}.
\newblock
\urldef\tempurl%
\url{https://doi.org/10.1037/a0022808}
\showDOI{\tempurl}


\bibitem[Ihle et~al\mbox{.}(2017)]%
        {ihle2017prospective}
\bibfield{author}{\bibinfo{person}{Andreas Ihle}, \bibinfo{person}{Paolo
  Ghisletta}, {and} \bibinfo{person}{Matthias Kliegel}.}
  \bibinfo{year}{2017}\natexlab{}.
\newblock \showarticletitle{Prospective memory and intraindividual variability
  in ongoing task response times in an adult lifespan sample: The role of cue
  focality}.
\newblock \bibinfo{journal}{\emph{Memory}} \bibinfo{volume}{25},
  \bibinfo{number}{3} (\bibinfo{year}{2017}), \bibinfo{pages}{370--376}.
\newblock
\urldef\tempurl%
\url{https://doi.org/10.1080/09658211.2016.1173705}
\showDOI{\tempurl}


\bibitem[Instagram(2020)]%
        {instagram_introducing_2020}
\bibfield{author}{\bibinfo{person}{Instagram}.}
  \bibinfo{year}{2020}\natexlab{}.
\newblock \bibinfo{title}{Introducing {Instagram} {Reels}}.
\newblock
\newblock
\urldef\tempurl%
\url{https://about.instagram.com/blog/announcements/introducing-instagram-reels-announcement}
\showURL{%
\tempurl}


\bibitem[Jaffe(2020)]%
        {jaffe_building_2020}
\bibfield{author}{\bibinfo{person}{Chris Jaffe}.}
  \bibinfo{year}{2020}\natexlab{}.
\newblock \bibinfo{title}{Building {YouTube} {Shorts}, a new way to watch \&
  create on {YouTube}}.
\newblock
\newblock
\urldef\tempurl%
\url{https://blog.youtube/news-and-events/building-youtube-shorts/}
\showURL{%
\tempurl}


\bibitem[Jager and Kliegel(2008)]%
        {jager2008time}
\bibfield{author}{\bibinfo{person}{Theodor Jager} {and}
  \bibinfo{person}{Matthias Kliegel}.} \bibinfo{year}{2008}\natexlab{}.
\newblock \showarticletitle{Time-based and event-based prospective memory
  across adulthood: Underlying mechanisms and differential costs on the ongoing
  task}.
\newblock \bibinfo{journal}{\emph{The Journal of general psychology}}
  \bibinfo{volume}{135}, \bibinfo{number}{1} (\bibinfo{year}{2008}),
  \bibinfo{pages}{4--22}.
\newblock
\urldef\tempurl%
\url{https://doi.org/10.3200/GENP.135.1.4-22}
\showDOI{\tempurl}


\bibitem[Kemp(2022)]%
        {kemp_digital_2022}
\bibfield{author}{\bibinfo{person}{Simon Kemp}.}
  \bibinfo{year}{2022}\natexlab{}.
\newblock \bibinfo{title}{Digital 2022: {Global} {Overview} {Report}}.
\newblock
\newblock
\urldef\tempurl%
\url{https://datareportal.com/reports/digital-2022-global-overview-report}
\showURL{%
\tempurl}


\bibitem[Kern et~al\mbox{.}(2010)]%
        {Kern2010GazemarksGV}
\bibfield{author}{\bibinfo{person}{Dagmar Kern}, \bibinfo{person}{Paul
  Marshall}, {and} \bibinfo{person}{Albrecht Schmidt}.}
  \bibinfo{year}{2010}\natexlab{}.
\newblock \showarticletitle{Gazemarks: gaze-based visual placeholders to ease
  attention switching}.
\newblock \bibinfo{journal}{\emph{Proceedings of the SIGCHI Conference on Human
  Factors in Computing Systems}} (\bibinfo{year}{2010}).
\newblock
\urldef\tempurl%
\url{https://doi.org/10.1145/1753326.1753646}
\showDOI{\tempurl}


\bibitem[Kvavilashvili and Fisher(2007)]%
        {kvavilashvili2007time}
\bibfield{author}{\bibinfo{person}{Lia Kvavilashvili} {and}
  \bibinfo{person}{Laura Fisher}.} \bibinfo{year}{2007}\natexlab{}.
\newblock \showarticletitle{Is time-based prospective remembering mediated by
  self-initiated rehearsals? Role of incidental cues, ongoing activity, age,
  and motivation.}
\newblock \bibinfo{journal}{\emph{Journal of Experimental Psychology: General}}
  \bibinfo{volume}{136}, \bibinfo{number}{1} (\bibinfo{year}{2007}),
  \bibinfo{pages}{112}.
\newblock
\urldef\tempurl%
\url{https://doi.org/10.1037/0096-3445.136.1.112}
\showDOI{\tempurl}


\bibitem[Lang et~al\mbox{.}(1999)]%
        {lang1999something}
\bibfield{author}{\bibinfo{person}{Annie Lang}, \bibinfo{person}{Robert~F
  Potter}, {and} \bibinfo{person}{Paul~D Bolls}.}
  \bibinfo{year}{1999}\natexlab{}.
\newblock \showarticletitle{Something for nothing: Is visual encoding
  automatic?}
\newblock \bibinfo{journal}{\emph{Media Psychology}} \bibinfo{volume}{1},
  \bibinfo{number}{2} (\bibinfo{year}{1999}), \bibinfo{pages}{145--163}.
\newblock
\urldef\tempurl%
\url{https://doi.org/10.1207/s1532785xmep0102_4}
\showDOI{\tempurl}


\bibitem[Langner et~al\mbox{.}(2010)]%
        {langner2010mental}
\bibfield{author}{\bibinfo{person}{Robert Langner}, \bibinfo{person}{Michael~B
  Steinborn}, \bibinfo{person}{Anjan Chatterjee}, \bibinfo{person}{Walter
  Sturm}, {and} \bibinfo{person}{Klaus Willmes}.}
  \bibinfo{year}{2010}\natexlab{}.
\newblock \showarticletitle{Mental fatigue and temporal preparation in simple
  reaction-time performance}.
\newblock \bibinfo{journal}{\emph{Acta psychologica}} \bibinfo{volume}{133},
  \bibinfo{number}{1} (\bibinfo{year}{2010}), \bibinfo{pages}{64--72}.
\newblock
\urldef\tempurl%
\url{https://doi.org/10.1016/j.actpsy.2009.10.001}
\showDOI{\tempurl}


\bibitem[Leroy(2009)]%
        {leroy2009so}
\bibfield{author}{\bibinfo{person}{Sophie Leroy}.}
  \bibinfo{year}{2009}\natexlab{}.
\newblock \showarticletitle{Why is it so hard to do my work? The challenge of
  attention residue when switching between work tasks}.
\newblock \bibinfo{journal}{\emph{Organizational Behavior and Human Decision
  Processes}} \bibinfo{volume}{109}, \bibinfo{number}{2}
  (\bibinfo{year}{2009}), \bibinfo{pages}{168--181}.
\newblock
\urldef\tempurl%
\url{https://doi.org/10.1016/j.obhdp.2009.04.002}
\showDOI{\tempurl}


\bibitem[Leroy and Glomb(2018)]%
        {leroy2018tasks}
\bibfield{author}{\bibinfo{person}{Sophie Leroy} {and}
  \bibinfo{person}{Theresa~M Glomb}.} \bibinfo{year}{2018}\natexlab{}.
\newblock \showarticletitle{Tasks interrupted: How anticipating time pressure
  on resumption of an interrupted task causes attention residue and low
  performance on interrupting tasks and how a “ready-to-resume” plan
  mitigates the effects}.
\newblock \bibinfo{journal}{\emph{Organization Science}} \bibinfo{volume}{29},
  \bibinfo{number}{3} (\bibinfo{year}{2018}), \bibinfo{pages}{380--397}.
\newblock
\urldef\tempurl%
\url{https://doi.org/10.1287/orsc.2017.1184}
\showDOI{\tempurl}


\bibitem[Leroy and Schmidt(2016)]%
        {leroy2016effect}
\bibfield{author}{\bibinfo{person}{Sophie Leroy} {and} \bibinfo{person}{Aaron~M
  Schmidt}.} \bibinfo{year}{2016}\natexlab{}.
\newblock \showarticletitle{The effect of regulatory focus on attention residue
  and performance during interruptions}.
\newblock \bibinfo{journal}{\emph{Organizational Behavior and Human Decision
  Processes}}  \bibinfo{volume}{137} (\bibinfo{year}{2016}),
  \bibinfo{pages}{218--235}.
\newblock
\urldef\tempurl%
\url{https://doi.org/10.1016/j.obhdp.2016.07.006}
\showDOI{\tempurl}


\bibitem[Lo and Andrews(2015)]%
        {lo2015transform}
\bibfield{author}{\bibinfo{person}{Steson Lo} {and} \bibinfo{person}{Sally
  Andrews}.} \bibinfo{year}{2015}\natexlab{}.
\newblock \showarticletitle{To transform or not to transform: Using generalized
  linear mixed models to analyse reaction time data}.
\newblock \bibinfo{journal}{\emph{Frontiers in psychology}}
  \bibinfo{volume}{6} (\bibinfo{year}{2015}), \bibinfo{pages}{1171}.
\newblock
\urldef\tempurl%
\url{https://doi.org/10.3389/fpsyg.2015.01171}
\showDOI{\tempurl}


\bibitem[Lu et~al\mbox{.}(2013)]%
        {lu2013supporting}
\bibfield{author}{\bibinfo{person}{Sara~A Lu}, \bibinfo{person}{Christopher~D
  Wickens}, \bibinfo{person}{Julie~C Prinet}, \bibinfo{person}{Shaun~D
  Hutchins}, \bibinfo{person}{Nadine Sarter}, {and} \bibinfo{person}{Angelia
  Sebok}.} \bibinfo{year}{2013}\natexlab{}.
\newblock \showarticletitle{Supporting interruption management and multimodal
  interface design: Three meta-analyses of task performance as a function of
  interrupting task modality}.
\newblock \bibinfo{journal}{\emph{Human factors}} \bibinfo{volume}{55},
  \bibinfo{number}{4} (\bibinfo{year}{2013}), \bibinfo{pages}{697--724}.
\newblock
\urldef\tempurl%
\url{https://doi.org/10.1177/001872081347629}
\showDOI{\tempurl}


\bibitem[Lukoff et~al\mbox{.}(2021)]%
        {lukoff2021design}
\bibfield{author}{\bibinfo{person}{Kai Lukoff}, \bibinfo{person}{Ulrik Lyngs},
  \bibinfo{person}{Himanshu Zade}, \bibinfo{person}{J~Vera Liao},
  \bibinfo{person}{James Choi}, \bibinfo{person}{Kaiyue Fan},
  \bibinfo{person}{Sean~A Munson}, {and} \bibinfo{person}{Alexis Hiniker}.}
  \bibinfo{year}{2021}\natexlab{}.
\newblock \showarticletitle{How the design of youtube influences user sense of
  agency}. In \bibinfo{booktitle}{\emph{Proceedings of the 2021 CHI Conference
  on Human Factors in Computing Systems}}. \bibinfo{pages}{1--17}.
\newblock
\urldef\tempurl%
\url{https://doi.org/10.1145/3411764.3445467}
\showDOI{\tempurl}


\bibitem[Lukoff et~al\mbox{.}(2018)]%
        {lukoff_what_2018}
\bibfield{author}{\bibinfo{person}{Kai Lukoff}, \bibinfo{person}{Cissy Yu},
  \bibinfo{person}{Julie Kientz}, {and} \bibinfo{person}{Alexis Hiniker}.}
  \bibinfo{year}{2018}\natexlab{}.
\newblock \showarticletitle{What {Makes} {Smartphone} {Use} {Meaningful} or
  {Meaningless}?}
\newblock  \bibinfo{volume}{2}, \bibinfo{number}{1} (\bibinfo{year}{2018}),
  \bibinfo{pages}{26}.
\newblock
\urldef\tempurl%
\url{https://doi.org/10.1145/3191754}
\showDOI{\tempurl}


\bibitem[Lupinacci(2020)]%
        {lupinacci_absentmindedly_2020}
\bibfield{author}{\bibinfo{person}{Ludmila Lupinacci}.}
  \bibinfo{year}{2020}\natexlab{}.
\newblock \showarticletitle{‘{Absentmindedly} scrolling through nothing’:
  liveness and compulsory continuous connectedness in social media}.
\newblock \bibinfo{journal}{\emph{Media, Culture \& Society}}
  (\bibinfo{date}{July} \bibinfo{year}{2020}),
  \bibinfo{pages}{0163443720939454}.
\newblock
\showISSN{0163-4437}
\urldef\tempurl%
\url{https://doi.org/10.1177/0163443720939454}
\showDOI{\tempurl}


\bibitem[Lyngs et~al\mbox{.}(2019)]%
        {lyngs2019self}
\bibfield{author}{\bibinfo{person}{Ulrik Lyngs}, \bibinfo{person}{Kai Lukoff},
  \bibinfo{person}{Petr Slovak}, \bibinfo{person}{Reuben Binns},
  \bibinfo{person}{Adam Slack}, \bibinfo{person}{Michael Inzlicht},
  \bibinfo{person}{Max Van~Kleek}, {and} \bibinfo{person}{Nigel Shadbolt}.}
  \bibinfo{year}{2019}\natexlab{}.
\newblock \showarticletitle{Self-control in cyberspace: Applying dual systems
  theory to a review of digital self-control tools}. In
  \bibinfo{booktitle}{\emph{proceedings of the 2019 CHI conference on human
  factors in computing systems}}. \bibinfo{pages}{1--18}.
\newblock
\urldef\tempurl%
\url{https://doi.org/10.1145/3290605.3300361}
\showDOI{\tempurl}


\bibitem[Mark et~al\mbox{.}(2008)]%
        {mark_cost_2008}
\bibfield{author}{\bibinfo{person}{Gloria Mark}, \bibinfo{person}{Daniela
  Gudith}, {and} \bibinfo{person}{Ulrich Klocke}.}
  \bibinfo{year}{2008}\natexlab{}.
\newblock \showarticletitle{The cost of interrupted work: {More} speed and
  stress}. \bibinfo{pages}{107--110}.
\newblock
\urldef\tempurl%
\url{https://doi.org/10.1145/1357054.1357072}
\showDOI{\tempurl}


\bibitem[Marty-Dugas et~al\mbox{.}(2018)]%
        {marty-dugas_relation_2018}
\bibfield{author}{\bibinfo{person}{Jeremy Marty-Dugas},
  \bibinfo{person}{Brandon Ralph}, \bibinfo{person}{Jonathan Oakman}, {and}
  \bibinfo{person}{Daniel Smilek}.} \bibinfo{year}{2018}\natexlab{}.
\newblock \showarticletitle{The {Relation} {Between} {Smartphone} {Use} and
  {Everyday} {Inattention}}.
\newblock \bibinfo{journal}{\emph{Psychology of Consciousness: Theory,
  Research, and Practice}}  \bibinfo{volume}{5} (\bibinfo{date}{March}
  \bibinfo{year}{2018}).
\newblock
\urldef\tempurl%
\url{https://doi.org/10.1037/cns0000131}
\showDOI{\tempurl}


\bibitem[McCoy(2016)]%
        {mccoy2016digital}
\bibfield{author}{\bibinfo{person}{Bernard~R McCoy}.}
  \bibinfo{year}{2016}\natexlab{}.
\newblock \showarticletitle{Digital distractions in the classroom phase II:
  Student classroom use of digital devices for non-class related purposes}.
\newblock  (\bibinfo{year}{2016}).
\newblock


\bibitem[McDaniel and Einstein(2007)]%
        {mcdaniel_prospective_2007}
\bibfield{author}{\bibinfo{person}{Mark~A. McDaniel} {and}
  \bibinfo{person}{Gilles~O. Einstein}.} \bibinfo{year}{2007}\natexlab{}.
\newblock \bibinfo{booktitle}{\emph{Prospective {Memory}: {An} {Overview} and
  {Synthesis} of an {Emerging} {Field}}}.
\newblock \bibinfo{publisher}{SAGE Publications}.
\newblock
\showISBNx{978-1-4833-1689-5}
\urldef\tempurl%
\url{https://doi.org/10.4135/9781452225913}
\showDOI{\tempurl}


\bibitem[McDaniel et~al\mbox{.}(2004a)]%
        {mcdaniel2004delaying}
\bibfield{author}{\bibinfo{person}{Mark~A McDaniel}, \bibinfo{person}{Gilles~O
  Einstein}, \bibinfo{person}{Thomas Graham}, {and} \bibinfo{person}{Erica
  Rall}.} \bibinfo{year}{2004}\natexlab{a}.
\newblock \showarticletitle{Delaying execution of intentions: Overcoming the
  costs of interruptions}.
\newblock \bibinfo{journal}{\emph{Applied Cognitive Psychology: The Official
  Journal of the Society for Applied Research in Memory and Cognition}}
  \bibinfo{volume}{18}, \bibinfo{number}{5} (\bibinfo{year}{2004}),
  \bibinfo{pages}{533--547}.
\newblock
\urldef\tempurl%
\url{https://doi.org/10.1002/acp.1002}
\showDOI{\tempurl}


\bibitem[McDaniel et~al\mbox{.}(2003)]%
        {mcdaniel2003aging}
\bibfield{author}{\bibinfo{person}{Mark~A McDaniel}, \bibinfo{person}{Gilles~O
  Einstein}, \bibinfo{person}{Amy~C Stout}, {and} \bibinfo{person}{Zack
  Morgan}.} \bibinfo{year}{2003}\natexlab{}.
\newblock \showarticletitle{Aging and maintaining intentions over delays: do it
  or lose it.}
\newblock \bibinfo{journal}{\emph{Psychology and Aging}} \bibinfo{volume}{18},
  \bibinfo{number}{4} (\bibinfo{year}{2003}), \bibinfo{pages}{823}.
\newblock
\urldef\tempurl%
\url{https://doi.org/10.1037/0882-7974.18.4.823}
\showDOI{\tempurl}


\bibitem[McDaniel et~al\mbox{.}(2004b)]%
        {mcdaniel_cue-focused_2004}
\bibfield{author}{\bibinfo{person}{Mark~A. McDaniel},
  \bibinfo{person}{Melissa~J. Guynn}, \bibinfo{person}{Gilles~O. Einstein},
  {and} \bibinfo{person}{Jennifer Breneiser}.}
  \bibinfo{year}{2004}\natexlab{b}.
\newblock \showarticletitle{Cue-{Focused} and {Reflexive}-{Associative}
  {Processes} in {Prospective} {Memory} {Retrieval}}.
\newblock \bibinfo{journal}{\emph{Journal of Experimental Psychology: Learning,
  Memory, and Cognition}} \bibinfo{volume}{30}, \bibinfo{number}{3}
  (\bibinfo{year}{2004}), \bibinfo{pages}{605--614}.
\newblock
\showISSN{1939-1285}
\urldef\tempurl%
\url{https://doi.org/10.1037/0278-7393.30.3.605}
\showDOI{\tempurl}


\bibitem[McFarlane and Latorella(2002)]%
        {mcfarlane2002scope}
\bibfield{author}{\bibinfo{person}{Daniel~C McFarlane} {and}
  \bibinfo{person}{Kara~A Latorella}.} \bibinfo{year}{2002}\natexlab{}.
\newblock \showarticletitle{The scope and importance of human interruption in
  human-computer interaction design}.
\newblock \bibinfo{journal}{\emph{Human-Computer Interaction}}
  \bibinfo{volume}{17}, \bibinfo{number}{1} (\bibinfo{year}{2002}),
  \bibinfo{pages}{1--61}.
\newblock
\urldef\tempurl%
\url{https://doi.org/10.1207/S15327051HCI1701_1}
\showDOI{\tempurl}


\bibitem[Mcgann et~al\mbox{.}(2002)]%
        {mcgann2002conceptual}
\bibfield{author}{\bibinfo{person}{Deborah Mcgann}, \bibinfo{person}{Judi~A
  Ellis}, {and} \bibinfo{person}{Alan Milne}.} \bibinfo{year}{2002}\natexlab{}.
\newblock \showarticletitle{Conceptual and perceptual processes in prospective
  remembering: Differential influence of attentional resources}.
\newblock \bibinfo{journal}{\emph{Memory \& cognition}} \bibinfo{volume}{30},
  \bibinfo{number}{7} (\bibinfo{year}{2002}), \bibinfo{pages}{1021--1032}.
\newblock
\urldef\tempurl%
\url{https://doi.org/10.3758/BF03194320}
\showDOI{\tempurl}


\bibitem[Min et~al\mbox{.}(2020)]%
        {min2020multimodal}
\bibfield{author}{\bibinfo{person}{Xiongkuo Min}, \bibinfo{person}{Guangtao
  Zhai}, \bibinfo{person}{Jiantao Zhou}, \bibinfo{person}{Xiao-Ping Zhang},
  \bibinfo{person}{Xiaokang Yang}, {and} \bibinfo{person}{Xinping Guan}.}
  \bibinfo{year}{2020}\natexlab{}.
\newblock \showarticletitle{A multimodal saliency model for videos with high
  audio-visual correspondence}.
\newblock \bibinfo{journal}{\emph{IEEE Transactions on Image Processing}}
  \bibinfo{volume}{29} (\bibinfo{year}{2020}), \bibinfo{pages}{3805--3819}.
\newblock
\urldef\tempurl%
\url{https://doi.org/10.1109/TIP.2020.2966082}
\showDOI{\tempurl}


\bibitem[Monacis et~al\mbox{.}(2017)]%
        {monacis_social_2017}
\bibfield{author}{\bibinfo{person}{Lucia Monacis}, \bibinfo{person}{Valeria~de
  Palo}, \bibinfo{person}{Mark~D. Griffiths}, {and} \bibinfo{person}{Maria
  Sinatra}.} \bibinfo{year}{2017}\natexlab{}.
\newblock \showarticletitle{Social networking addiction, attachment style, and
  validation of the {Italian} version of the {Bergen} {Social} {Media}
  {Addiction} {Scale}}.
\newblock \bibinfo{journal}{\emph{Journal of Behavioral Addictions}}
  \bibinfo{volume}{6}, \bibinfo{number}{2} (\bibinfo{date}{June}
  \bibinfo{year}{2017}), \bibinfo{pages}{178--186}.
\newblock
\showISSN{2063-5303, 2062-5871}
\urldef\tempurl%
\url{https://doi.org/10.1556/2006.6.2017.023}
\showDOI{\tempurl}


\bibitem[Monk and Kidd(2008)]%
        {monk2008effects}
\bibfield{author}{\bibinfo{person}{Christopher~A Monk} {and}
  \bibinfo{person}{David~G Kidd}.} \bibinfo{year}{2008}\natexlab{}.
\newblock \showarticletitle{The effects of brief interruptions on task
  resumption}. In \bibinfo{booktitle}{\emph{Proceedings of the Human Factors
  and Ergonomics Society Annual Meeting}}, Vol.~\bibinfo{volume}{52}. SAGE
  Publications Sage CA: Los Angeles, CA, \bibinfo{pages}{403--407}.
\newblock
\urldef\tempurl%
\url{https://doi.org/10.1177/154193120805200443}
\showDOI{\tempurl}


\bibitem[Monk et~al\mbox{.}(2008)]%
        {monk2008effect}
\bibfield{author}{\bibinfo{person}{Christopher~A Monk},
  \bibinfo{person}{J~Gregory Trafton}, {and} \bibinfo{person}{Deborah~A
  Boehm-Davis}.} \bibinfo{year}{2008}\natexlab{}.
\newblock \showarticletitle{The effect of interruption duration and demand on
  resuming suspended goals.}
\newblock \bibinfo{journal}{\emph{Journal of experimental psychology: Applied}}
  \bibinfo{volume}{14}, \bibinfo{number}{4} (\bibinfo{year}{2008}),
  \bibinfo{pages}{299}.
\newblock
\urldef\tempurl%
\url{https://doi.org/10.1037/a0014402}
\showDOI{\tempurl}


\bibitem[Montag et~al\mbox{.}(2021)]%
        {montag2021psychology}
\bibfield{author}{\bibinfo{person}{Christian Montag}, \bibinfo{person}{Haibo
  Yang}, {and} \bibinfo{person}{Jon~D Elhai}.} \bibinfo{year}{2021}\natexlab{}.
\newblock \showarticletitle{On the psychology of TikTok use: A first glimpse
  from empirical findings}.
\newblock \bibinfo{journal}{\emph{Frontiers in public health}}
  \bibinfo{volume}{9} (\bibinfo{year}{2021}), \bibinfo{pages}{641673}.
\newblock
\urldef\tempurl%
\url{https://doi.org/10.3389/fpubh.2021.641673}
\showDOI{\tempurl}


\bibitem[Morgan et~al\mbox{.}(2009)]%
        {morgan2009improving}
\bibfield{author}{\bibinfo{person}{Phillip~L Morgan}, \bibinfo{person}{John
  Patrick}, \bibinfo{person}{Samuel~M Waldron}, \bibinfo{person}{Sophia~L
  King}, {and} \bibinfo{person}{Tanya Patrick}.}
  \bibinfo{year}{2009}\natexlab{}.
\newblock \showarticletitle{Improving memory after interruption: Exploiting
  soft constraints and manipulating information access cost.}
\newblock \bibinfo{journal}{\emph{Journal of experimental psychology: Applied}}
  \bibinfo{volume}{15}, \bibinfo{number}{4} (\bibinfo{year}{2009}),
  \bibinfo{pages}{291}.
\newblock
\urldef\tempurl%
\url{https://doi.org/10.1037/a0018008}
\showDOI{\tempurl}


\bibitem[Nees and Fortna(2015)]%
        {nees2015comparison}
\bibfield{author}{\bibinfo{person}{Michael~A Nees} {and}
  \bibinfo{person}{Anjali Fortna}.} \bibinfo{year}{2015}\natexlab{}.
\newblock \showarticletitle{A comparison of human versus virtual
  interruptions}.
\newblock \bibinfo{journal}{\emph{Ergonomics}} \bibinfo{volume}{58},
  \bibinfo{number}{5} (\bibinfo{year}{2015}), \bibinfo{pages}{852--856}.
\newblock
\urldef\tempurl%
\url{https://doi.org/10.1080/00140139.2014.990934}
\showDOI{\tempurl}


\bibitem[O'Conaill and Frohlich(1995)]%
        {o1995timespace}
\bibfield{author}{\bibinfo{person}{Brid O'Conaill} {and} \bibinfo{person}{David
  Frohlich}.} \bibinfo{year}{1995}\natexlab{}.
\newblock \showarticletitle{Timespace in the workplace: Dealing with
  interruptions}. In \bibinfo{booktitle}{\emph{Conference companion on Human
  factors in computing systems}}. \bibinfo{pages}{262--263}.
\newblock
\urldef\tempurl%
\url{https://doi.org/10.1145/223355.223665}
\showDOI{\tempurl}


\bibitem[Oldham et~al\mbox{.}(1991)]%
        {oldham1991physical}
\bibfield{author}{\bibinfo{person}{Greg~R Oldham}, \bibinfo{person}{Carol~T
  Kulik}, {and} \bibinfo{person}{Lee~P Stepina}.}
  \bibinfo{year}{1991}\natexlab{}.
\newblock \showarticletitle{Physical environments and employee reactions:
  effects of stimulus-screening skills and job complexity}.
\newblock \bibinfo{journal}{\emph{Academy of Management Journal}}
  \bibinfo{volume}{34}, \bibinfo{number}{4} (\bibinfo{year}{1991}),
  \bibinfo{pages}{929--938}.
\newblock
\urldef\tempurl%
\url{https://doi.org/10.5465/256397}
\showDOI{\tempurl}


\bibitem[Peirce(2007)]%
        {peirce2007psychopy}
\bibfield{author}{\bibinfo{person}{Jonathan~W Peirce}.}
  \bibinfo{year}{2007}\natexlab{}.
\newblock \showarticletitle{PsychoPy—psychophysics software in Python}.
\newblock \bibinfo{journal}{\emph{Journal of neuroscience methods}}
  \bibinfo{volume}{162}, \bibinfo{number}{1-2} (\bibinfo{year}{2007}),
  \bibinfo{pages}{8--13}.
\newblock
\urldef\tempurl%
\url{https://doi.org/10.1016/j.jneumeth.2006.11.017}
\showDOI{\tempurl}


\bibitem[Peng and Lu(2012)]%
        {peng2012model}
\bibfield{author}{\bibinfo{person}{Heng Peng} {and} \bibinfo{person}{Ying Lu}.}
  \bibinfo{year}{2012}\natexlab{}.
\newblock \showarticletitle{Model selection in linear mixed effect models}.
\newblock \bibinfo{journal}{\emph{Journal of Multivariate Analysis}}
  \bibinfo{volume}{109} (\bibinfo{year}{2012}), \bibinfo{pages}{109--129}.
\newblock
\showISSN{0047-259X}
\urldef\tempurl%
\url{https://doi.org/10.1016/j.jmva.2012.02.005}
\showDOI{\tempurl}


\bibitem[Powell et~al\mbox{.}(2019)]%
        {powell2019framing}
\bibfield{author}{\bibinfo{person}{Thomas~E Powell}, \bibinfo{person}{Hajo~G
  Boomgaarden}, \bibinfo{person}{Knut De~Swert}, {and} \bibinfo{person}{Claes~H
  de Vreese}.} \bibinfo{year}{2019}\natexlab{}.
\newblock \showarticletitle{Framing fast and slow: A dual processing account of
  multimodal framing effects}.
\newblock \bibinfo{journal}{\emph{Media Psychology}} \bibinfo{volume}{22},
  \bibinfo{number}{4} (\bibinfo{year}{2019}), \bibinfo{pages}{572--600}.
\newblock
\urldef\tempurl%
\url{https://doi.org/10.1080/15213269.2018.1476891}
\showDOI{\tempurl}


\bibitem[Ratcliff(1978)]%
        {ratcliff1978ddm}
\bibfield{author}{\bibinfo{person}{Roger Ratcliff}.}
  \bibinfo{year}{1978}\natexlab{}.
\newblock \showarticletitle{A theory of memory retrieval}.
\newblock \bibinfo{journal}{\emph{Psychological Review}}  \bibinfo{volume}{85}
  (\bibinfo{year}{1978}), \bibinfo{pages}{59–108}.
\newblock
\showISSN{1939-1471}
\urldef\tempurl%
\url{https://doi.org/10.1037/0033-295X.85.2.59}
\showDOI{\tempurl}


\bibitem[Ratcliff(2013)]%
        {ratcliff2013parameter}
\bibfield{author}{\bibinfo{person}{Roger Ratcliff}.}
  \bibinfo{year}{2013}\natexlab{}.
\newblock \showarticletitle{Parameter variability and distributional
  assumptions in the diffusion model.}
\newblock \bibinfo{journal}{\emph{Psychological review}} \bibinfo{volume}{120},
  \bibinfo{number}{1} (\bibinfo{year}{2013}), \bibinfo{pages}{281}.
\newblock
\urldef\tempurl%
\url{https://doi.org/10.1037/a0030775}
\showDOI{\tempurl}


\bibitem[Razali et~al\mbox{.}(2011)]%
        {razali2011power}
\bibfield{author}{\bibinfo{person}{Nornadiah~Mohd Razali},
  \bibinfo{person}{Yap~Bee Wah}, {et~al\mbox{.}}}
  \bibinfo{year}{2011}\natexlab{}.
\newblock \showarticletitle{Power comparisons of shapiro-wilk,
  kolmogorov-smirnov, lilliefors and anderson-darling tests}.
\newblock \bibinfo{journal}{\emph{Journal of statistical modeling and
  analytics}} \bibinfo{volume}{2}, \bibinfo{number}{1} (\bibinfo{year}{2011}),
  \bibinfo{pages}{21--33}.
\newblock


\bibitem[Richter and Yeung(2012)]%
        {richter_memory_2012}
\bibfield{author}{\bibinfo{person}{Franziska~R. Richter} {and}
  \bibinfo{person}{Nick Yeung}.} \bibinfo{year}{2012}\natexlab{}.
\newblock \showarticletitle{Memory and {Cognitive} {Control} in {Task}
  {Switching}}.
\newblock \bibinfo{journal}{\emph{Psychological Science}} \bibinfo{volume}{23},
  \bibinfo{number}{10} (\bibinfo{date}{Oct.} \bibinfo{year}{2012}),
  \bibinfo{pages}{1256--1263}.
\newblock
\showISSN{0956-7976}
\urldef\tempurl%
\url{https://doi.org/10.1177/0956797612444613}
\showDOI{\tempurl}
\newblock
\shownote{Publisher: SAGE Publications Inc}.


\bibitem[Rogers(2014)]%
        {rogers_level_2014}
\bibfield{author}{\bibinfo{person}{Scott Rogers}.}
  \bibinfo{year}{2014}\natexlab{}.
\newblock \bibinfo{booktitle}{\emph{Level {Up}! {The} {Guide} to {Great}
  {Video} {Game} {Design}}}.
\newblock \bibinfo{publisher}{John Wiley \& Sons}.
\newblock
\showISBNx{978-1-118-87719-7}
\newblock
\shownote{Google-Books-ID: UT5jAwAAQBAJ}.


\bibitem[Rummel et~al\mbox{.}(2013)]%
        {rummel2013performance}
\bibfield{author}{\bibinfo{person}{Jan Rummel}, \bibinfo{person}{Beatrice~G
  Kuhlmann}, {and} \bibinfo{person}{Dayna~R Touron}.}
  \bibinfo{year}{2013}\natexlab{}.
\newblock \showarticletitle{Performance predictions affect attentional
  processes of event-based prospective memory}.
\newblock \bibinfo{journal}{\emph{Consciousness and cognition}}
  \bibinfo{volume}{22}, \bibinfo{number}{3} (\bibinfo{year}{2013}),
  \bibinfo{pages}{729--741}.
\newblock
\urldef\tempurl%
\url{https://doi.org/10.1016/j.concog.2013.04.012}
\showDOI{\tempurl}


\bibitem[Rummel et~al\mbox{.}(2017)]%
        {rummel2017role}
\bibfield{author}{\bibinfo{person}{Jan Rummel}, \bibinfo{person}{Ann-Katrin
  Wesslein}, {and} \bibinfo{person}{Thorsten Meiser}.}
  \bibinfo{year}{2017}\natexlab{}.
\newblock \showarticletitle{The role of action coordination for prospective
  memory: Task-interruption demands affect intention realization.}
\newblock \bibinfo{journal}{\emph{Journal of Experimental Psychology: Learning,
  Memory, and Cognition}} \bibinfo{volume}{43}, \bibinfo{number}{5}
  (\bibinfo{year}{2017}), \bibinfo{pages}{717}.
\newblock
\urldef\tempurl%
\url{https://doi.org/10.1037/xlm0000334}
\showDOI{\tempurl}


\bibitem[Sanderson and Grundgeiger(2015)]%
        {sanderson2015interruptions}
\bibfield{author}{\bibinfo{person}{Penelope~M Sanderson} {and}
  \bibinfo{person}{Tobias Grundgeiger}.} \bibinfo{year}{2015}\natexlab{}.
\newblock \showarticletitle{How do interruptions affect clinician performance
  in healthcare? Negotiating fidelity, control, and potential generalizability
  in the search for answers}.
\newblock \bibinfo{journal}{\emph{International Journal of Human-Computer
  Studies}}  \bibinfo{volume}{79} (\bibinfo{year}{2015}),
  \bibinfo{pages}{85--96}.
\newblock
\urldef\tempurl%
\url{https://doi.org/10.1016/j.ijhcs.2014.11.003}
\showDOI{\tempurl}


\bibitem[Sanjram and Azizuddin(2010)]%
        {sanjram2010attention}
\bibfield{author}{\bibinfo{person}{Premjit~K Sanjram} {and}
  \bibinfo{person}{Khan Azizuddin}.} \bibinfo{year}{2010}\natexlab{}.
\newblock \showarticletitle{Attention and programmer characteristics in
  prospective memory: an investigation of habit intrusion error in programmer
  multitasking}. In \bibinfo{booktitle}{\emph{Proceedings of the 28th Annual
  European Conference on Cognitive Ergonomics}}. \bibinfo{pages}{307--310}.
\newblock
\urldef\tempurl%
\url{https://doi.org/10.1145/1962300.1962365}
\showDOI{\tempurl}


\bibitem[Sasangohar et~al\mbox{.}(2012)]%
        {sasangohar2012not}
\bibfield{author}{\bibinfo{person}{Farzan Sasangohar}, \bibinfo{person}{Birsen
  Donmez}, \bibinfo{person}{Patricia Trbovich}, {and}
  \bibinfo{person}{Anthony~C Easty}.} \bibinfo{year}{2012}\natexlab{}.
\newblock \showarticletitle{Not all interruptions are created equal: positive
  interruptions in healthcare}. In \bibinfo{booktitle}{\emph{Proceedings of the
  Human Factors and Ergonomics Society Annual Meeting}},
  Vol.~\bibinfo{volume}{56}. SAGE Publications Sage CA: Los Angeles, CA,
  \bibinfo{pages}{824--828}.
\newblock
\urldef\tempurl%
\url{https://doi.org/10.1177/10711813125611}
\showDOI{\tempurl}


\bibitem[Schacter(2012)]%
        {schacter_constructive_2012}
\bibfield{author}{\bibinfo{person}{Daniel~L. Schacter}.}
  \bibinfo{year}{2012}\natexlab{}.
\newblock \showarticletitle{Constructive memory: past and future}.
\newblock \bibinfo{journal}{\emph{Dialogues in Clinical Neuroscience}}
  \bibinfo{volume}{14}, \bibinfo{number}{1} (\bibinfo{date}{March}
  \bibinfo{year}{2012}), \bibinfo{pages}{7--18}.
\newblock
\showISSN{null}
\urldef\tempurl%
\url{https://doi.org/10.31887/DCNS.2012.14.1/dschacter}
\showDOI{\tempurl}


\bibitem[Schulze(2003)]%
        {schulze_memos_2003}
\bibfield{author}{\bibinfo{person}{Hendrik Schulze}.}
  \bibinfo{year}{2003}\natexlab{}.
\newblock \showarticletitle{{MEMOS}: an interactive assistive system for
  prospective memory deficit compensation-architecture and functionality}. In
  \bibinfo{booktitle}{\emph{Proceedings of the 6th international {ACM}
  {SIGACCESS} conference on {Computers} and accessibility}}
  \emph{(\bibinfo{series}{Assets '04})}. \bibinfo{publisher}{Association for
  Computing Machinery}, \bibinfo{address}{New York, NY, USA},
  \bibinfo{pages}{79--85}.
\newblock
\showISBNx{978-1-58113-911-2}
\urldef\tempurl%
\url{https://doi.org/10.1145/1028630.1028645}
\showDOI{\tempurl}


\bibitem[Scullin et~al\mbox{.}(2010)]%
        {scullin_focalnonfocal_2010}
\bibfield{author}{\bibinfo{person}{Michael~K. Scullin},
  \bibinfo{person}{Mark~A. McDaniel}, \bibinfo{person}{Jill~T. Shelton}, {and}
  \bibinfo{person}{Ji~Hae Lee}.} \bibinfo{year}{2010}\natexlab{}.
\newblock \showarticletitle{Focal/nonfocal cue effects in prospective memory:
  {Monitoring} difficulty or different retrieval processes?}
\newblock \bibinfo{journal}{\emph{Journal of Experimental Psychology: Learning,
  Memory, and Cognition}} \bibinfo{volume}{36}, \bibinfo{number}{3}
  (\bibinfo{year}{2010}), \bibinfo{pages}{736--749}.
\newblock
\showISSN{1939-1285}
\urldef\tempurl%
\url{https://doi.org/10.1037/a0018971}
\showDOI{\tempurl}


\bibitem[Sharifian and Zahodne(2021)]%
        {sharifian2021daily}
\bibfield{author}{\bibinfo{person}{Neika Sharifian} {and}
  \bibinfo{person}{Laura~B Zahodne}.} \bibinfo{year}{2021}\natexlab{}.
\newblock \showarticletitle{Daily associations between social media use and
  memory failures: The mediating role of negative affect}.
\newblock \bibinfo{journal}{\emph{The Journal of general psychology}}
  \bibinfo{volume}{148}, \bibinfo{number}{1} (\bibinfo{year}{2021}),
  \bibinfo{pages}{67--83}.
\newblock
\urldef\tempurl%
\url{https://doi.org/10.1080/00221309.2020.1743228}
\showDOI{\tempurl}


\bibitem[Shelton et~al\mbox{.}(2019)]%
        {shelton2019multiprocess}
\bibfield{author}{\bibinfo{person}{Jill~Talley Shelton},
  \bibinfo{person}{Michael~K Scullin}, {and} \bibinfo{person}{Jessica~Y
  Hacker}.} \bibinfo{year}{2019}\natexlab{}.
\newblock \showarticletitle{The multiprocess framework: Historical context and
  the “dynamic” extension}.
\newblock In \bibinfo{booktitle}{\emph{Prospective memory}}.
  \bibinfo{publisher}{Routledge}, \bibinfo{pages}{10--26}.
\newblock
\urldef\tempurl%
\url{https://doi.org/10.4324/9781351000154}
\showDOI{\tempurl}


\bibitem[Shinn et~al\mbox{.}(2020)]%
        {shinn2020flexible}
\bibfield{author}{\bibinfo{person}{Maxwell Shinn}, \bibinfo{person}{Norman~H
  Lam}, {and} \bibinfo{person}{John~D Murray}.}
  \bibinfo{year}{2020}\natexlab{}.
\newblock \showarticletitle{A flexible framework for simulating and fitting
  generalized drift-diffusion models}.
\newblock \bibinfo{journal}{\emph{ELife}}  \bibinfo{volume}{9}
  (\bibinfo{year}{2020}), \bibinfo{pages}{e56938}.
\newblock
\urldef\tempurl%
\url{https://doi.org/10.7554/eLife.56938}
\showDOI{\tempurl}


\bibitem[Sindermann et~al\mbox{.}(2020)]%
        {sindermann2020predicting}
\bibfield{author}{\bibinfo{person}{Cornelia Sindermann}, \bibinfo{person}{Jon~D
  Elhai}, {and} \bibinfo{person}{Christian Montag}.}
  \bibinfo{year}{2020}\natexlab{}.
\newblock \showarticletitle{Predicting tendencies towards the disordered use of
  Facebook's social media platforms: On the role of personality, impulsivity,
  and social anxiety}.
\newblock \bibinfo{journal}{\emph{Psychiatry Research}}  \bibinfo{volume}{285}
  (\bibinfo{year}{2020}), \bibinfo{pages}{112793}.
\newblock
\urldef\tempurl%
\url{https://doi.org/10.1016/j.psychres.2020.112793}
\showDOI{\tempurl}


\bibitem[Smith(2003)]%
        {smith2003cost}
\bibfield{author}{\bibinfo{person}{Rebekah~E Smith}.}
  \bibinfo{year}{2003}\natexlab{}.
\newblock \showarticletitle{The cost of remembering to remember in event-based
  prospective memory: investigating the capacity demands of delayed intention
  performance.}
\newblock \bibinfo{journal}{\emph{Journal of Experimental Psychology: Learning,
  Memory, and Cognition}} \bibinfo{volume}{29}, \bibinfo{number}{3}
  (\bibinfo{year}{2003}), \bibinfo{pages}{347}.
\newblock
\urldef\tempurl%
\url{https://doi.org/10.1037/0278-7393.29.3.347}
\showDOI{\tempurl}


\bibitem[Smith(2008)]%
        {smith2008connecting}
\bibfield{author}{\bibinfo{person}{Rebekah~E Smith}.}
  \bibinfo{year}{2008}\natexlab{}.
\newblock \showarticletitle{Connecting the past and the future: Attention,
  memory, and delayed intentions.}
\newblock  (\bibinfo{year}{2008}).
\newblock


\bibitem[Smith(2010)]%
        {smith2010costs}
\bibfield{author}{\bibinfo{person}{Rebekah~E Smith}.}
  \bibinfo{year}{2010}\natexlab{}.
\newblock \showarticletitle{What costs do reveal and moving beyond the cost
  debate: Reply to Einstein and McDaniel (2010).}
\newblock  (\bibinfo{year}{2010}).
\newblock
\urldef\tempurl%
\url{https://doi.org/10.1037/a0019183}
\showDOI{\tempurl}


\bibitem[Smith et~al\mbox{.}(2012)]%
        {smith2012prospective}
\bibfield{author}{\bibinfo{person}{Rebekah~E Smith},
  \bibinfo{person}{Sebastian~S Horn}, {and} \bibinfo{person}{Ute~J Bayen}.}
  \bibinfo{year}{2012}\natexlab{}.
\newblock \showarticletitle{Prospective memory in young and older adults: The
  effects of ongoing-task load}.
\newblock \bibinfo{journal}{\emph{Aging, Neuropsychology, and Cognition}}
  \bibinfo{volume}{19}, \bibinfo{number}{4} (\bibinfo{year}{2012}),
  \bibinfo{pages}{495--514}.
\newblock
\urldef\tempurl%
\url{https://doi.org/10.1080/13825585.2013.827150}
\showDOI{\tempurl}


\bibitem[Smith et~al\mbox{.}(2007)]%
        {smith2007cost}
\bibfield{author}{\bibinfo{person}{Rebekah~E Smith}, \bibinfo{person}{R~Reed
  Hunt}, \bibinfo{person}{Jennifer~C McVay}, {and} \bibinfo{person}{Melissa~D
  McConnell}.} \bibinfo{year}{2007}\natexlab{}.
\newblock \showarticletitle{The cost of event-based prospective memory: salient
  target events.}
\newblock \bibinfo{journal}{\emph{Journal of Experimental Psychology: Learning,
  Memory, and Cognition}} \bibinfo{volume}{33}, \bibinfo{number}{4}
  (\bibinfo{year}{2007}), \bibinfo{pages}{734}.
\newblock
\urldef\tempurl%
\url{https://doi.org/10.1037/0278-7393.33.4.734}
\showDOI{\tempurl}


\bibitem[Smith and Loft(2014)]%
        {smith2014investigating}
\bibfield{author}{\bibinfo{person}{Rebekah~E Smith} {and}
  \bibinfo{person}{Shayne Loft}.} \bibinfo{year}{2014}\natexlab{}.
\newblock \showarticletitle{Investigating the cost to ongoing tasks not
  associated with prospective memory task requirements}.
\newblock \bibinfo{journal}{\emph{Consciousness and cognition}}
  \bibinfo{volume}{27} (\bibinfo{year}{2014}), \bibinfo{pages}{1--13}.
\newblock


\bibitem[Spira and Feintuch(2005)]%
        {spira_cost_2005}
\bibfield{author}{\bibinfo{person}{J.~B. Spira} {and} \bibinfo{person}{J.~B.
  Feintuch}.} \bibinfo{year}{2005}\natexlab{}.
\newblock \bibinfo{booktitle}{\emph{The cost of not paying attention: {How}
  interruptions impact knowledge worker productivity. {New} {York}, {NY}:
  {Basex}}}.
\newblock \bibinfo{type}{{T}echnical {R}eport}. \bibinfo{institution}{Basex
  Inc}, \bibinfo{address}{New York, NY}. \bibinfo{pages}{21} pages.
\newblock
\urldef\tempurl%
\url{http://iorgforum.org/wp-content/uploads/2011/06/CostOfNotPayingAttention.BasexReport1.pdf}
\showURL{%
\tempurl}


\bibitem[Strickland et~al\mbox{.}(2019)]%
        {strickland2019evidence}
\bibfield{author}{\bibinfo{person}{Luke Strickland}, \bibinfo{person}{Shayne
  Loft}, {and} \bibinfo{person}{Andrew Heathcote}.}
  \bibinfo{year}{2019}\natexlab{}.
\newblock \showarticletitle{Evidence accumulation modeling of event-based
  prospective memory}.
\newblock In \bibinfo{booktitle}{\emph{Prospective Memory}}.
  \bibinfo{publisher}{Routledge}, \bibinfo{pages}{78--94}.
\newblock


\bibitem[Terzimehić et~al\mbox{.}(2022)]%
        {terzimehic_mindphone_2022}
\bibfield{author}{\bibinfo{person}{Nađa Terzimehić}, \bibinfo{person}{Luke
  Haliburton}, \bibinfo{person}{Philipp Greiner}, \bibinfo{person}{Albrecht
  Schmidt}, \bibinfo{person}{Heinrich Hussmann}, {and} \bibinfo{person}{Ville
  Mäkelä}.} \bibinfo{year}{2022}\natexlab{}.
\newblock \showarticletitle{{MindPhone}: {Mindful} {Reflection} at {Unlock}
  {Can} {Reduce} {Absentminded} {Smartphone} {Use}}. In
  \bibinfo{booktitle}{\emph{Designing {Interactive} {Systems} {Conference}}}
  \emph{(\bibinfo{series}{{DIS} '22})}. \bibinfo{publisher}{Association for
  Computing Machinery}, \bibinfo{address}{New York, NY, USA},
  \bibinfo{pages}{1818--1830}.
\newblock
\showISBNx{978-1-4503-9358-4}
\urldef\tempurl%
\url{https://doi.org/10.1145/3532106.3533575}
\showDOI{\tempurl}


\bibitem[Thorson and Lang(1988)]%
        {thorson1988effects}
\bibfield{author}{\bibinfo{person}{Esther Thorson} {and} \bibinfo{person}{Annie
  Lang}.} \bibinfo{year}{1988}\natexlab{}.
\newblock \showarticletitle{The effects of videographic complexity on memory
  for televised information}.
\newblock \bibinfo{journal}{\emph{International Communication Association, New
  Orleans, LA}} (\bibinfo{year}{1988}).
\newblock


\bibitem[Uncapher et~al\mbox{.}(2017)]%
        {uncapher2017media}
\bibfield{author}{\bibinfo{person}{Melina~R Uncapher}, \bibinfo{person}{Lin
  Lin}, \bibinfo{person}{Larry~D Rosen}, \bibinfo{person}{Heather~L Kirkorian},
  \bibinfo{person}{Naomi~S Baron}, \bibinfo{person}{Kira Bailey},
  \bibinfo{person}{Joanne Cantor}, \bibinfo{person}{David~L Strayer},
  \bibinfo{person}{Thomas~D Parsons}, {and} \bibinfo{person}{Anthony~D
  Wagner}.} \bibinfo{year}{2017}\natexlab{}.
\newblock \showarticletitle{Media multitasking and cognitive, psychological,
  neural, and learning differences}.
\newblock \bibinfo{journal}{\emph{Pediatrics}} \bibinfo{volume}{140},
  \bibinfo{number}{Supplement\_2} (\bibinfo{year}{2017}),
  \bibinfo{pages}{S62--S66}.
\newblock
\urldef\tempurl%
\url{https://doi.org/10.1542/peds.2016-1758D}
\showDOI{\tempurl}


\bibitem[van Ravenzwaaij et~al\mbox{.}(2019)]%
        {van2019bayes}
\bibfield{author}{\bibinfo{person}{Don van Ravenzwaaij}, \bibinfo{person}{Rei
  Monden}, \bibinfo{person}{Jorge~N Tendeiro}, {and} \bibinfo{person}{John
  Ioannidis}.} \bibinfo{year}{2019}\natexlab{}.
\newblock \showarticletitle{Bayes factors for superiority, non-inferiority, and
  equivalence designs}.
\newblock \bibinfo{journal}{\emph{BMC medical research methodology}}
  \bibinfo{volume}{19}, \bibinfo{number}{1} (\bibinfo{year}{2019}),
  \bibinfo{pages}{1--12}.
\newblock
\urldef\tempurl%
\url{https://doi.org/10.1186/s12874-019-0699-7}
\showDOI{\tempurl}


\bibitem[Vinding et~al\mbox{.}(2021)]%
        {vinding2021volition}
\bibfield{author}{\bibinfo{person}{Mikkel~C. Vinding},
  \bibinfo{person}{Jonas~Kristoffer Lindeløv}, \bibinfo{person}{Yahui Xiao},
  \bibinfo{person}{Raymond C.~K. Chan}, {and} \bibinfo{person}{Thomas~Alrik
  Sørensen}.} \bibinfo{year}{2021}\natexlab{}.
\newblock \showarticletitle{Volition in prospective Memory: Evidence against
  differences between free and fixed target events}.
\newblock \bibinfo{journal}{\emph{Consciousness and Cognition}}
  \bibinfo{volume}{94} (\bibinfo{date}{Sep} \bibinfo{year}{2021}),
  \bibinfo{pages}{103175}.
\newblock
\showISSN{1053-8100}
\urldef\tempurl%
\url{https://doi.org/10.1016/j.concog.2021.103175}
\showDOI{\tempurl}


\bibitem[Voss and Voss(2008)]%
        {voss2008fast}
\bibfield{author}{\bibinfo{person}{Andreas Voss} {and} \bibinfo{person}{Jochen
  Voss}.} \bibinfo{year}{2008}\natexlab{}.
\newblock \showarticletitle{A fast numerical algorithm for the estimation of
  diffusion model parameters}.
\newblock \bibinfo{journal}{\emph{Journal of Mathematical Psychology}}
  \bibinfo{volume}{52}, \bibinfo{number}{1} (\bibinfo{year}{2008}),
  \bibinfo{pages}{1--9}.
\newblock
\urldef\tempurl%
\url{https://doi.org/10.1016/j.jmp.2007.09.005}
\showDOI{\tempurl}


\bibitem[Wagenmakers et~al\mbox{.}(2008)]%
        {wagenmakers2008diffusion}
\bibfield{author}{\bibinfo{person}{Eric-Jan Wagenmakers},
  \bibinfo{person}{Roger Ratcliff}, \bibinfo{person}{Pablo Gomez}, {and}
  \bibinfo{person}{Gail McKoon}.} \bibinfo{year}{2008}\natexlab{}.
\newblock \showarticletitle{A diffusion model account of criterion shifts in
  the lexical decision task}.
\newblock \bibinfo{journal}{\emph{Journal of memory and language}}
  \bibinfo{volume}{58}, \bibinfo{number}{1} (\bibinfo{year}{2008}),
  \bibinfo{pages}{140--159}.
\newblock
\urldef\tempurl%
\url{https://doi.org/10.1016/j.jml.2007.04.006}
\showDOI{\tempurl}


\bibitem[Wang and Pérez-Quiñones(2014)]%
        {wang_exploring_2014}
\bibfield{author}{\bibinfo{person}{Yao Wang} {and} \bibinfo{person}{Manuel~A.
  Pérez-Quiñones}.} \bibinfo{year}{2014}\natexlab{}.
\newblock \showarticletitle{Exploring the role of prospective memory in
  location-based reminders}. In \bibinfo{booktitle}{\emph{Proceedings of the
  2014 {ACM} {International} {Joint} {Conference} on {Pervasive} and
  {Ubiquitous} {Computing}: {Adjunct} {Publication}}}
  \emph{(\bibinfo{series}{{UbiComp} '14 {Adjunct}})}.
  \bibinfo{publisher}{Association for Computing Machinery},
  \bibinfo{address}{New York, NY, USA}, \bibinfo{pages}{1373--1380}.
\newblock
\showISBNx{978-1-4503-3047-3}
\urldef\tempurl%
\url{https://doi.org/10.1145/2638728.2641718}
\showDOI{\tempurl}


\bibitem[Wilmer et~al\mbox{.}(2017)]%
        {wilmer2017smartphones}
\bibfield{author}{\bibinfo{person}{Henry~H Wilmer}, \bibinfo{person}{Lauren~E
  Sherman}, {and} \bibinfo{person}{Jason~M Chein}.}
  \bibinfo{year}{2017}\natexlab{}.
\newblock \showarticletitle{Smartphones and cognition: A review of research
  exploring the links between mobile technology habits and cognitive
  functioning}.
\newblock \bibinfo{journal}{\emph{Frontiers in psychology}}
  \bibinfo{volume}{8} (\bibinfo{year}{2017}), \bibinfo{pages}{605}.
\newblock
\urldef\tempurl%
\url{https://doi.org/10.3389/fpsyg.2017.00605}
\showDOI{\tempurl}


\bibitem[Wirz and Schwabe(2020)]%
        {wirz2020prioritized}
\bibfield{author}{\bibinfo{person}{Lisa Wirz} {and} \bibinfo{person}{Lars
  Schwabe}.} \bibinfo{year}{2020}\natexlab{}.
\newblock \showarticletitle{Prioritized attentional processing: Acute stress,
  memory and stimulus emotionality facilitate attentional disengagement}.
\newblock \bibinfo{journal}{\emph{Neuropsychologia}}  \bibinfo{volume}{138}
  (\bibinfo{year}{2020}), \bibinfo{pages}{107334}.
\newblock
\urldef\tempurl%
\url{https://doi.org/10.1016/j.neuropsychologia.2020.107334}
\showDOI{\tempurl}


\bibitem[Wobbrock et~al\mbox{.}(2011)]%
        {wobbrock2011art}
\bibfield{author}{\bibinfo{person}{Jacob~O. Wobbrock}, \bibinfo{person}{Leah
  Findlater}, \bibinfo{person}{Darren Gergle}, {and} \bibinfo{person}{James~J.
  Higgins}.} \bibinfo{year}{2011}\natexlab{}.
\newblock \showarticletitle{The Aligned Rank Transform for Nonparametric
  Factorial Analyses Using Only Anova Procedures}. In
  \bibinfo{booktitle}{\emph{Proceedings of the SIGCHI Conference on Human
  Factors in Computing Systems}} (Vancouver, BC, Canada)
  \emph{(\bibinfo{series}{CHI '11})}. \bibinfo{publisher}{Association for
  Computing Machinery}, \bibinfo{address}{New York, NY, USA},
  \bibinfo{pages}{143–146}.
\newblock
\showISBNx{9781450302289}
\urldef\tempurl%
\url{https://doi.org/10.1145/1978942.1978963}
\showDOI{\tempurl}


\bibitem[Yorke-Smith et~al\mbox{.}(2009)]%
        {yorke2009like}
\bibfield{author}{\bibinfo{person}{Neil Yorke-Smith}, \bibinfo{person}{Shahin
  Saadati}, \bibinfo{person}{Karen~L Myers}, {and} \bibinfo{person}{David~N
  Morley}.} \bibinfo{year}{2009}\natexlab{}.
\newblock \showarticletitle{Like an intuitive and courteous butler: A proactive
  personal agent for task management}. In \bibinfo{booktitle}{\emph{Proceedings
  of The 8th International Conference on Autonomous Agents and Multiagent
  Systems-Volume 1}}. \bibinfo{pages}{337--344}.
\newblock
\urldef\tempurl%
\url{https://doi.org/10.5555/1558013.1558059}
\showDOI{\tempurl}


\bibitem[Zeng et~al\mbox{.}(2021)]%
        {zeng2021research}
\bibfield{author}{\bibinfo{person}{Jing Zeng}, \bibinfo{person}{Chrystal
  Abidin}, {and} \bibinfo{person}{Mike~S Sch{\"a}fer}.}
  \bibinfo{year}{2021}\natexlab{}.
\newblock \showarticletitle{Research perspectives on TikTok and its legacy
  apps: introduction}.
\newblock \bibinfo{journal}{\emph{International Journal of Communication}}
  \bibinfo{volume}{15} (\bibinfo{year}{2021}), \bibinfo{pages}{3161--3172}.
\newblock
\urldef\tempurl%
\url{https://doi.org/10.5167/uzh-205427}
\showDOI{\tempurl}


\bibitem[Zhang et~al\mbox{.}(2022)]%
        {zhang2022heavy}
\bibfield{author}{\bibinfo{person}{Jie Zhang}, \bibinfo{person}{Tongtong Xue},
  \bibinfo{person}{Shaobo Liu}, {and} \bibinfo{person}{Zhijie Zhang}.}
  \bibinfo{year}{2022}\natexlab{}.
\newblock \showarticletitle{Heavy and light media multitaskers employ different
  neurocognitive strategies in a prospective memory task: An ERP study}.
\newblock \bibinfo{journal}{\emph{Computers in Human Behavior}}
  \bibinfo{volume}{135} (\bibinfo{year}{2022}), \bibinfo{pages}{107379}.
\newblock
\urldef\tempurl%
\url{https://doi.org/10.1016/j.chb.2022.107379}
\showDOI{\tempurl}


\bibitem[Zheng(2021)]%
        {zheng_influence_2021}
\bibfield{author}{\bibinfo{person}{Mengyu Zheng}.}
  \bibinfo{year}{2021}\natexlab{}.
\newblock \showarticletitle{Influence of {Short} {Video} {Watching} {Behaviors}
  on {Visual} {Short}-{Term} {Memory}}. \bibinfo{publisher}{Atlantis Press},
  \bibinfo{pages}{1855--1859}.
\newblock
\showISBNx{978-94-6239-495-7}
\urldef\tempurl%
\url{https://doi.org/10.2991/assehr.k.211220.314}
\showDOI{\tempurl}


\end{thebibliography}

\end{document}